\documentclass{aa}
\usepackage{graphicx}
\usepackage{amsmath,amssymb}
\usepackage{times}

\newcommand{\Bozic}{{Bo\v{z}i\'{c}}}
\newcommand{\Zdarsky}{{\v{Z}\v{d}\'arsk\'{y}}}

\newcommand{\kms}{\,\ensuremath{\mathrm{km}~\mathrm{s}^{-1}}}
\newcommand{\Halpha}{\ensuremath{\mathrm{H}\alpha}}
\newcommand{\Hbeta}{\ensuremath{\mathrm{H}\beta}}
\newcommand{\Hgamma}{\ensuremath{\mathrm{H}\gamma}}
\newcommand{\Hdelta}{\ensuremath{\mathrm{H}\delta}}
\newcommand{\WHalpha}{\ensuremath{W_{\mathrm{H}\alpha}}}

\newcommand{\Teff}{\ensuremath{T_\mathrm{eff}}}
\newcommand{\K}{\,\ensuremath{\mathrm K}}
\newcommand{\kdra}{\mbox{\object{$\varkappa$~Dra}}}
\newcommand{\Delnup}{\ensuremath{\Delta\nu_\mathrm{p}}}
\newcommand{\ms}{\ensuremath{\mathrm{M}_{\sun}}}
\newcommand{\rs}{\ensuremath{\mathrm{R}_{\sun}}}
\newcommand{\bv}{\ensuremath{B\!-\!V}}
\newcommand{\ub}{\ensuremath{U\!-\!B}}
\newcommand{\ubv}{\ensuremath{U\!BV}}

\newcommand{\SPEFO}{{\tt SPEFO}}
\newcommand{\Reticon}{{\em Reticon}}
\newcommand{\HEROS}{{\em HEROS}}
\newcommand{\CCD}{{\em CCD}}

\begin{document}

\title{Properties and nature of Be stars\thanks{Tables 2 to 4 are
only available in electronic form at CDS
({\tt http://cdsweb.u-strasbg.fr/cats/J.A+A.htx}) and at
{\tt http://www.edpsciences.org}.}}
\subtitle{XXIII. Long-term variations and physical properties of
       \object{\kdra}}
\titlerunning{Long-term variations and physical properties of the Be
star \object{\kdra}}
\author{S.~M.~Saad\inst{1,2}\and
J.~Kub\'{a}t\inst{1}\and
P.~Koubsk\'y\inst{1}\and
P.~Harmanec\inst{3,1}\and
P.~\v{S}koda\inst{1}\and
D.~Kor\v{c}\'akov\'a\inst{1}\and
J.~Krti\v{c}ka\inst{4,1}\and
M.~\v{S}lechta\inst{1}\and
H.~Bo\v{z}i\'c\inst{5}\and
H.~Ak\inst{6}\and
P.~Hadrava\inst{1}\and
V.~Votruba\inst{4,1}
}
\authorrunning{S. M. Saad et al.}

\offprints{J.~Kub\'at, \\
\email{kubat@sunstel.asu.cas.cz}}

\institute{
Astronomick\'y \'ustav, Akademie v\v{e}d \v{C}esk\'e
republiky, CZ-251 65 Ond\v{r}ejov, Czech Republic
\and
National Research Institute of Astronomy and Geophysics,
11421 Helwan, Cairo, Egypt
\and
Astronomick\'y \'ustav UK, V Hole\v{s}ovi\v{c}k\'ach 2,
CZ-180 00 Praha 8, Czech Republic
\and
\'Ustav teoretick\'e fyziky a astrofyziky P\v{r}F MU,
Kotl\'a\v{r}sk\'a 2, CZ-611 37 Brno, Czech Republic
\and
Opservatorij Hvar, Geodetski fakultet, Sveu\v{c}ili\v{s}te Zagreb,
10000 Zagreb, Croatia
\and
Ankara University, Science Faculty, Astronomy and Space Science Dept.,
Tando\u{g}an, Ankara, 06100 Turkey
}
\date{Received 1 September 2003 / Accepted 7 January 2004}

\abstract{
We present an analysis of new spectroscopic observations of the
bright Be star {\kdra} obtained at the Ond\v{r}ejov observatory during
1992 -- 2003 and {\ubv} photometric observations secured at several
observatories.
General characteristics and a line identification of the spectrum of
{\kdra} are obtained in the regions 3730 -- 5650\,{\AA} and 5850 --
7800\,{\AA} by a comparison with the theoretical spectrum.
The fundamental stellar parameters have been obtained from a comparison
with a grid of NLTE model atmospheres.
The best fit was found for ${\Teff}=14\,000\K$, $\log g = 3.5$, and
$v \sin i = 170\kms$.
These values together with a Hipparcos parallax lead to a stellar mass
$M=4.8\pm0.8\,\ms$ and radius $R=6.4\pm0.5\,\rs$.
It is encouraging to see that these values agree well with the expected
evolutionary mass and radius for the effective temperature we derived.
Long-term variations of {\kdra} were analysed using measurements of
equivalent widths, central intensities, peak intensities of emission
lines and emission peak velocity differences for {\Halpha}, {\Hbeta},
{\Hgamma}, {\Hdelta}, and some helium, silicon, and iron lines.
It turned out that the previously reported period of 23 years in the
variation of the emission strength is probably a cyclic, not a strictly
periodic phenomenon. An attempt to find out a period from all available
records of the {\Hbeta} emission strength led to a value of
(8044 $\pm$ 167)~days (22.0~years) but the phase plots show
that each cycle has a different shape and length.
The maximum strength of the emission {\sl lags} behind the brightness
maximum.
This is a behaviour usually observed for long-term changes of Be stars
with a positive correlation between the brightness and emission
strength.
Since there are obviously no published speckle observations of the star,
we suggest these should be carried out.
They could help to deny or confirm the possibility that the emission
episodes are triggered by a periastron passage of a putative binary
companion moving in an eccentric orbit with a 8044-d period, as it seems
to be the case for some Be binaries.
For the moment, the nature and origin of the disk around {\kdra} still
remains unknown.
From the comparison of the electronic spectra obtained at different
phases of the long-term cycle and synthetic spectra it appears that
there are no detectable changes in the photospheric part of the Balmer
lines related to variations in the Balmer emission strength which could
be attributed to an extended photosphere corresponding to inner parts of
the disk, optically thick in continuum.

}
\maketitle

\section{Introduction}

{\kdra} (5~Dra, HD\,109387, HR\,4787, BD +70\degr703, MWC\,222,
HIP\,61281) is one of the brightest variable ($V=3\fm75$ -- $3\fm95$)
Northern Hemisphere Be stars, for which the bright {\Halpha} line had
already been reported by Campbell (\cite{wahid}).
Since that, {\kdra} has been a subject of numerous investigations, with
many controversial reports on the time scales and character of its
variability in various observables.
This made it one of the most challenging objects for further studies.
A detailed history of its investigation is summarized in papers by
Juza et al. (\cite{Juza91}, \cite{Juza94}) and Hirata (\cite{talatabe})
and need not be repeated here.

The most important results of several more recent studies can be
summarized as follows:
\begin{enumerate}
\item Periodic variations of the {\Hbeta} emission strength with a
      period of 23 years, discovered by Jessup (\cite{J32}), have been
      confirmed and detected also in photometry and continuum
      polarimetry.
      Their period was improved to 8406 days (23.01 years; Juza et al.
      \cite{Juza94}).
\item The maximum brightness in the long-term cycle precedes the maximum
      of the emission strength for a few years (Juza et al.
      \cite{Juza94}).
      The maximum of the emission strength coincides with the maximum of
      continuum polarization (Arsenijevi\'c et al. \cite{france94}).
      These facts seem to indicate that the brightening of the object is
      related to temporarily extended photosphere which can be
      satisfactorily modelled by rotating model photospheres (Hirata
      \cite{talatabe}).
\item {\kdra} is the primary component of a single-line spectroscopic
      binary in a circular orbit with an orbital period of $61\fd55$ and
      a semi-amplitude of $K\sim 7-8\kms$.
      This orbital radial-velocity (RV hereafter) variations could be
      detected in virtually all published RVs.
      They are accompanied by parallel phase-locked $V/R$ variations of
      the double Balmer emission lines (Juza et al. \cite{Juza91}).
\item A controversial period of rapid variations of $0\fd890384$ or its
      various aliases could be found in all available RV data sets,
      line asymmetry and polarization and its value corresponds well to
      the expected rotational period of the Be primary (Juza et al.
      \cite{Juza91}).
\item The rapid variations in the form of line width and asymmetry and
      also travelling sub-features, seen in a series of electronic
      spectra, can best be reconciled with a period of $0\fd545$ (one of
      the aliases of the $0\fd89$ period according to Juza et al.
      \cite{Juza91}) and can be interpreted as signatures of nonradial
      pulsations (Hill \cite{Hill91}).
\end{enumerate}

It is clear that understanding the long-term variability is one of the
most important clues to understanding the nature of the still mysterious
Be phenomenon.
Such a task requires, among other things, a very patient collection of
observational data and their careful reduction.
Since {\kdra} has been on our observing program both at Ond\v{r}ejov
(spectroscopy with electronic detectors), and at Hvar, and recently also
at Tubitak ({\ubv} photometry) for more than a decade, it was deemed
useful to publish these unique data sets in extenso, with relevant
analyses, to enable further comparisons with current and future
theoretical models of the Be phenomenon.
We also offer some interpretations and derive improved basic physical
properties of the Be star based on detailed comparison with synthetic
line profiles over a large range of wavelengths.

\section{Observations and data reduction}
\subsection{Spectroscopy}

\begin{table*}
\caption{The observational journal of {\kdra} at the Ond\v{r}ejov
Observatory.}\label{obsjour}
\begin{tabular}{ccclccc}
\hline\noalign{\smallskip}
\textbf{Epoch}&
\textbf{No. of}&
\textbf{Spectrograph}&
\textbf{Detector}&
\textbf{Resolving}&
\textbf{Spectral} \\
{[HJD$-$2400000]}&
\bf spectra&
&
&
\bf power&
\bf range [\AA] \\
\noalign{\smallskip}\hline\noalign{\smallskip}
48000 -- 51714 & 93
& coud\'{e} & {\Reticon} RL-1872F/30 & 10\,000 & 6300 -- 6740 \\
49021 -- 49026 & 3
& coud\'{e} & {\Reticon} RL-1872F/30 & 10\,000 & 4310 -- 4520 \\
49079 -- 49116 & 6
& coud\'{e} & {\Reticon} RL-1872F/30 & 10\,000 & 4750 -- 4960 \\
52321 -- 52321 & 1
&coud\'{e} & {\CCD} SITe005 800$\times$2000& 10\,000 & 4300 -- 4554 \\
52323 -- 52323 & 2
&coud\'{e} & {\CCD} SITe005 800$\times$2000& 10\,000& 6256 -- 6769 \\
51900 -- 52727 & 30
& \HEROS & {\CCD} EEV 2000$\times$800 &20\,000& 3450 -- 5650 \\
& & & {\CCD} EEV 1152$\times$770 & 20\,000 & 5850 -- 8620 \\
52742 -- 52754 &6
&coud\'{e} & {\CCD} SITe005 800$\times$2000& 10\,000 & 4753 -- 5005  \\
52734 -- 52754 & 22
&coud\'{e} & {\CCD} SITe005 800$\times$2000& 10\,000 & 6256 -- 6769 \\
\noalign{\smallskip}\hline
\end{tabular}
\end{table*}

Electronic spectra of {\kdra} were secured from June 1992 to April 2003
in the coud\'e and Cassegrain foci of the Ond\v{r}ejov 2m telescope.
The details of observations are summarized in Table \ref{obsjour}.
\begin{itemize}
\item Most of the spectra (102) were obtained in the coud\'e focus with
      a {\em Reticon RL-1872F/30} detector at a reciprocal dispersion of
      17\,{\AA}/mm;
      93 of them cover the {\Halpha} region, 6 of them contain {\Hbeta}
      and 3 the {\Hgamma} line.
\item Since January 2001 to March 2003, 30 spectra were obtained
      at the red and blue channels of the fibre-fed echelle
      spectrograph {\HEROS} ({\em H}eidelberg {\em E}xtended
      {\em R}ange {\em O}ptical {\em S}pectrograph) attached to the
      Cassegrain focus. See, e.g., Stahl et al. (\cite{jad}),
      \v{S}tefl \& Rivinius (\cite{SR00}) or
      \v{S}koda \& \v{S}lechta (\cite{herpopis}) for more details
      about {\HEROS}.
\item 31 spectra were obtained with a {\CCD} camera attached to
      the coud\'{e} focus; 24 of them near {\Halpha}, 6 near {\Hbeta}
      and one around {\Hgamma}.
\end{itemize}

A complete reduction of {\Reticon} spectra, i.e. wavelength calibration,
rectification, and intensity and equivalent-width ($W$) measurements
were all carried out using the program {\SPEFO} written by Dr.\,J.\,Horn
(see Horn et al. \cite{H92}, \v{S}koda 1996 for details).
All initial reduction of {\HEROS} and {\CCD} spectra (bias subtraction,
flat-fielding and wavelength calibration) were carried out using
modified {\tt MIDAS} and {\tt IRAF} packages, respectively.
The unrectified, wavelength-calibrated spectra were then exported into
{\SPEFO} where the rectification and all line parameter measurements
were done.

\subsection{Photometry}

A rich collection of {\ubv} photometric observations was obtained at
Hvar and Skalnat\'e Pleso, and complemented by observations collected
from the astronomical literature, by Juza et al. (\cite{Juza94}).
In addition to these observations, we obtained new {\ubv} photometry
of {\kdra} at two observatories:
10 observations at Hvar and
47 observations at Tubitak.
All these observations were corrected for differential extinction and
transformed into a standard {\ubv} system (based on the mean values
given by Johnson et al. \cite{joh66}) via non-linear transformation
formul\ae\ using {\tt HEC22/VYPAR} photometric software (see Harmanec
\& Horn \cite{hec22} and Harmanec et al. \cite{homo} for the details and
copies of the programs).
We also extracted 115 $H_p$ photometric observations having photometric
flag 0 or 1 from the Hipparcos archive (Perryman et al. {\cite{hip}) and
transformed them into Johnson $V$ magnitude using Harmanec's
(\cite{hec98b}) formula.
Finally, we read out 96 $V$ magnitude differences from a graph published
in a poster paper by Croxall et al. (\cite{karen}) and shifted them
properly to get them on a scale of other data using the nights where
this data set overlaps with our calibrated data.

A few comments are relevant:
\begin{enumerate}
\item To ensure the best homogeneity of Hvar data, we only used
      differential photometry of {\kdra} obtained relative to
      \object{HR\,4687} = \object{HD\,107193} for which the following
      all-sky {\ubv} values were adopted on the basis of recent data
      homogenization
      $V = 5\fm489$, $\bv = +0\fm038$, $\ub = +0\fm079$.
      The check star \object{HR\,5018} = \object{HD\,115612}
      ($V = 6\fm215$, $\bv = -0\fm056$, $\ub = -0\fm137$)
      was observed as frequently as the variable.
\item The same comparisons were also used at the 0.4-m reflector at
      Tubitak.
\item All older Hvar and Skalnat\'e Pleso {\ubv} observations, published
      by Juza et al. (\cite{Juza94}), were also re-reduced relative to
      the above values for the comparison.
\end{enumerate}

\section{The optical spectrum of \kdra}

The vast majority of the new spectra cover the wavelength region from
about 6300 to 6700\,{\AA}, which includes the {\Halpha} line.
This spectral region is characterized by a {\em permanent} presence of
emission in {\Halpha} and \ion{Fe}{ii} 6456\,{\AA} lines, while the
lines \ion{Si}{ii} 6347\,{\AA}, \ion{Si}{ii} 6371\,{\AA}, and
\ion{He}{i} 6678\,{\AA} are seen in absorption.
In a much wider spectral range 3450 -- 8620\,{\AA}, covered only by
the {\HEROS} spectra, the Balmer lines up to H15 as well as some
infrared lines are seen.

The general time behaviour documented by the spectral region around
{\Halpha} seems to be confirmed by the rest of the visible spectrum.
As usually, the Balmer emission is strongest for {\Halpha} and
decreases toward higher members of the series.
A number of \ion{Fe}{ii} emission lines are present in the whole optical
spectrum, whereas both the triplet and singlet \ion{He}{i} lines are in
absorption.
Lines of other ions, like \ion{Mg}{ii}, \ion{C}{ii}, \ion{O}{i}
are all in absorption, with a notable exception of the infrared
oxygen triplet \ion{O}{i} 7772, 7774, 7775 {\AA}.
Note that certain parts of the red and infrared spectrum are strongly
contaminated by the telluric absorption lines and bands.

\subsection{Comparison with a theoretical spectrum}\label{modcomp}

\begin{figure*}
\resizebox{\hsize}{!}{\includegraphics{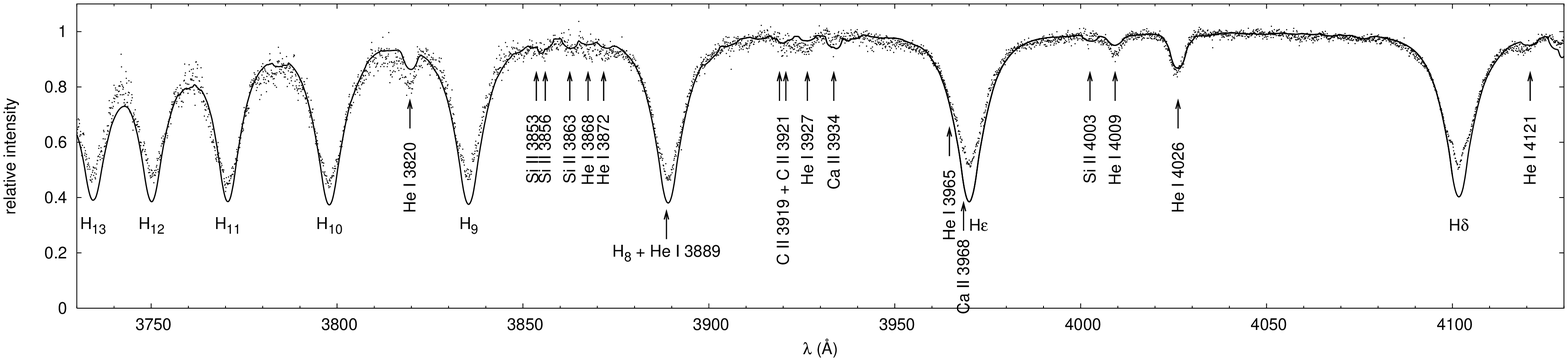}}
\resizebox{\hsize}{!}{\includegraphics{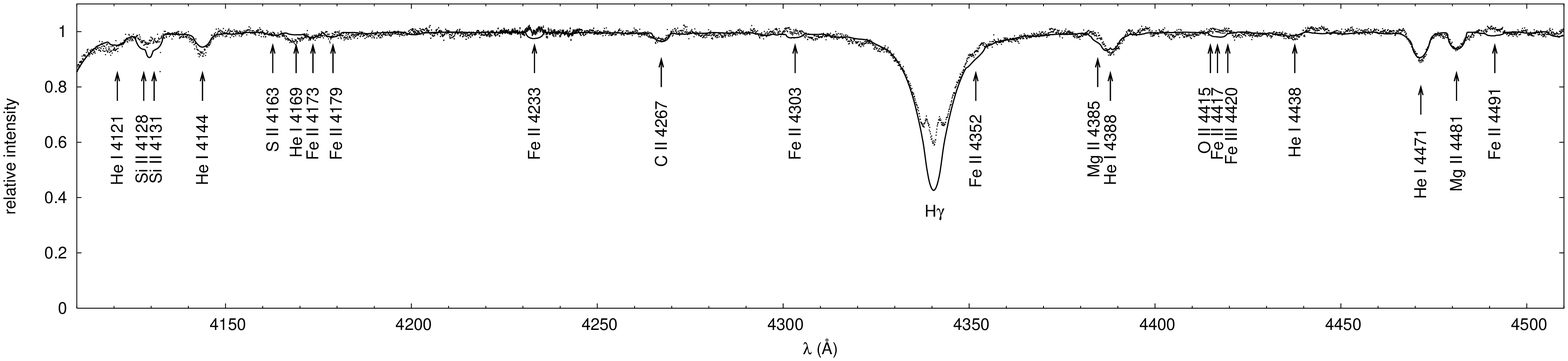}}
\resizebox{\hsize}{!}{\includegraphics{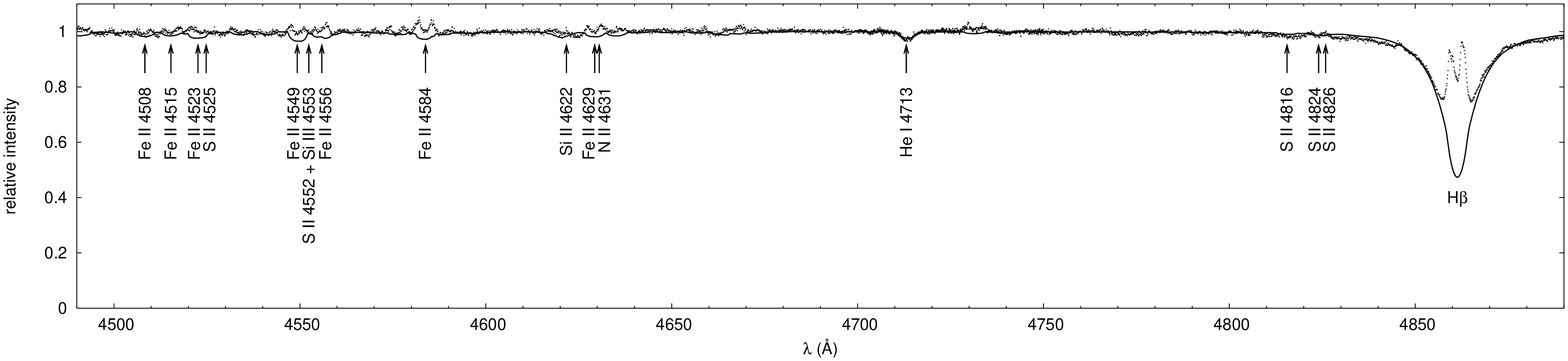}}
\resizebox{\hsize}{!}{\includegraphics{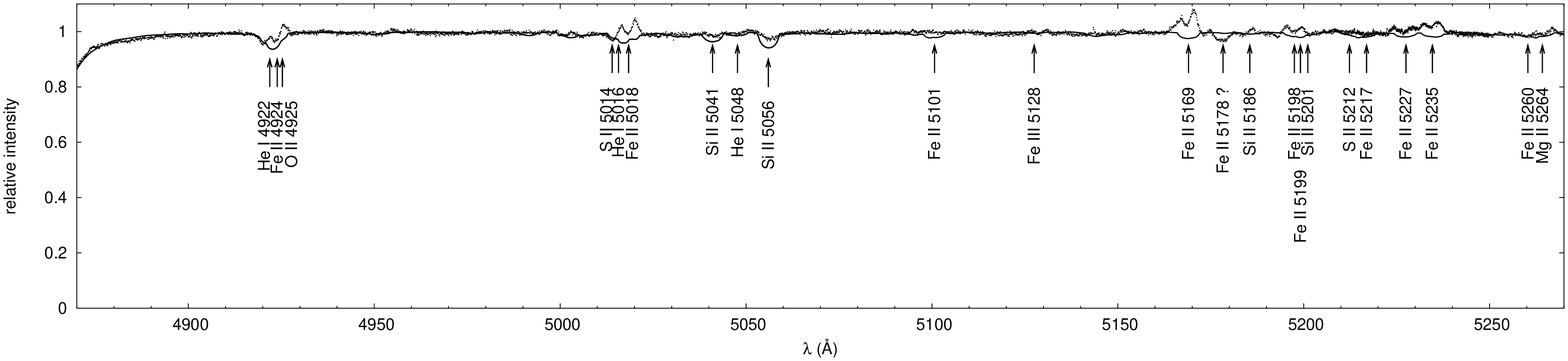}}
\resizebox{\hsize}{!}{\includegraphics{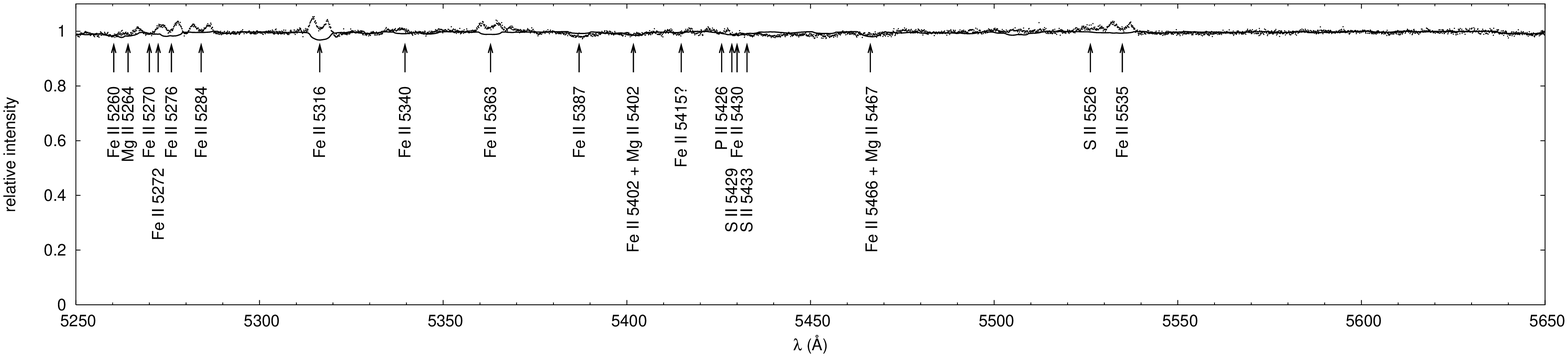}}
\caption{Line identification and comparison of the spectrum of {\kdra}
obtained using the blue channel of the {\HEROS} spectrograph (dots) and
the synthetic NLTE spectrum $\Teff=14\,000\K$, $\log g=3.5$ (full line).}
\label{HERblue}
\end{figure*}

\begin{figure*}
\resizebox{\hsize}{!}{\includegraphics{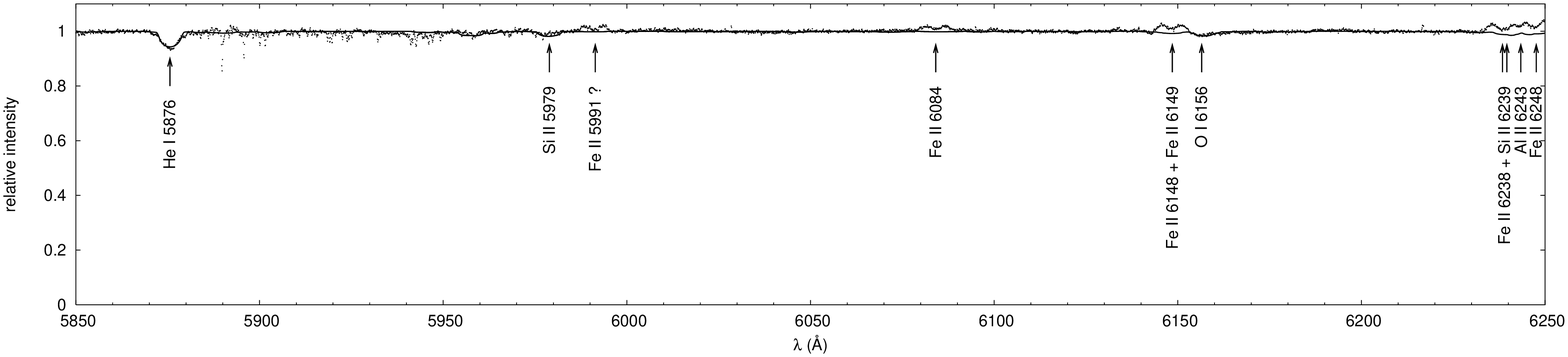}}
\resizebox{\hsize}{!}{\includegraphics{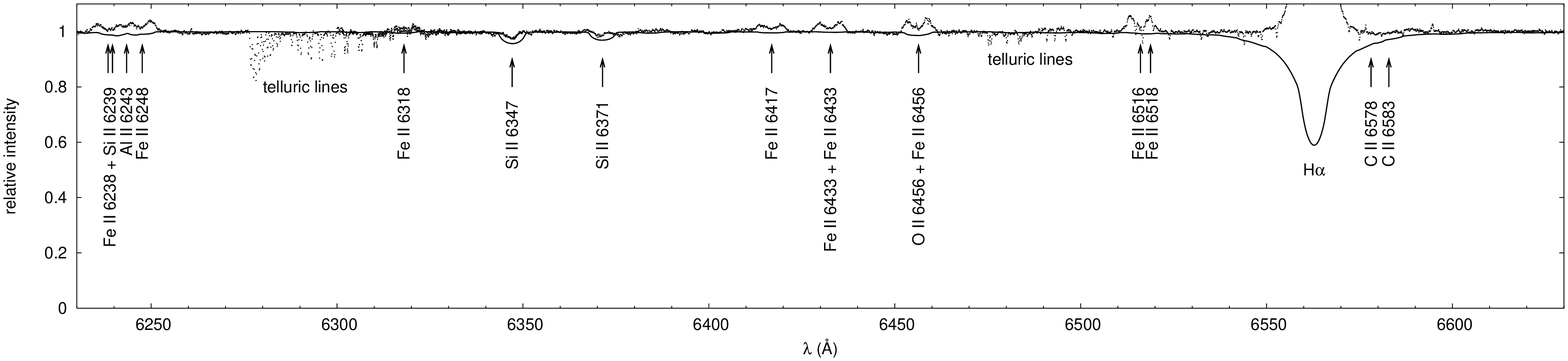}}
\resizebox{\hsize}{!}{\includegraphics{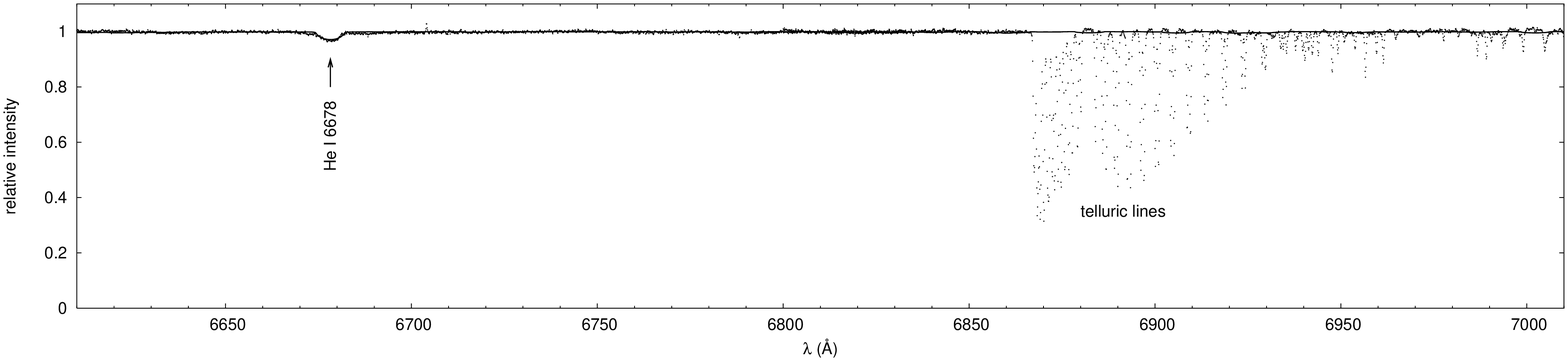}}
\resizebox{\hsize}{!}{\includegraphics{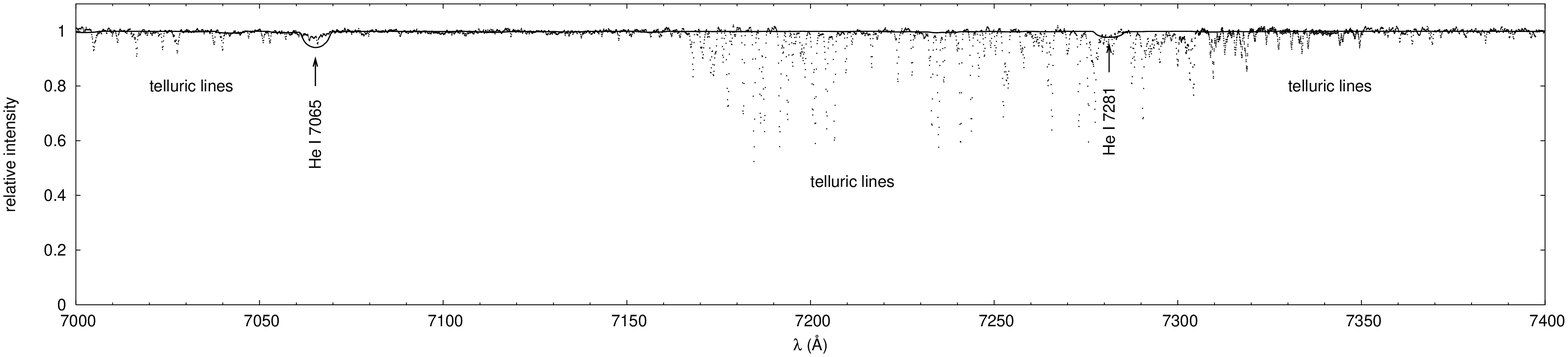}}
\resizebox{\hsize}{!}{\includegraphics{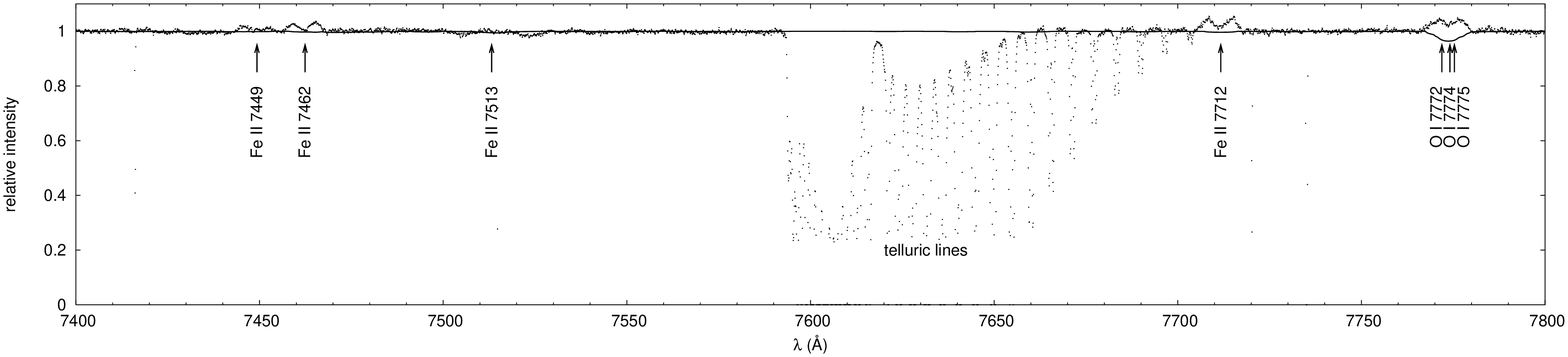}}
\caption{Line identification and comparison of the spectrum of {\kdra}
obtained using the red channel of the {\HEROS} spectrograph (dots) and
the synthetic NLTE spectrum $\Teff=14\,000\K$, $\log g=3.5$ (full line).
Note the presence of telluric lines and bands near $6280${\AA},
{\Halpha}, $6870${\AA}, $7250${\AA}, and $7590${\AA.}}
\label{HERred}
\end{figure*}

Determination of basic parameters of stellar atmospheres (the effective
temperature {\Teff} and the surface gravity $g$) is a relatively
straightforward procedure for normal, main sequence stars, where
simplifying assumptions of local thermodynamic equilibrium (LTE) in a
static atmosphere and plane-parallel geometry are fulfilled with a
sufficient accuracy.

However, this is not the case for Be stars.
The very presence of the emission lines indicates the violation of the
assumptions of a plane-parallel geometry and a static atmosphere.
Yet, one can try to derive {\em approximate} values based on standard
models, keeping their limited applicability in mind.
This should work rather well in circumstances when the star in question
is temporarily without any observable signatures of the Be envelope and
has a pure absorption spectrum.
Indeed, very satisfactory results for two other Be stars, \object{4~Her}
(\object{HD\,142926}) and \object{60~Cyg} (\object{HD\,200310}), were
obtained this way -- cf. Koubsk\'y et al. (\cite{4Her}) and Koubsk\'y
et al. (\cite{60Cyg}), respectively.

Very regrettably, {\kdra} has never been observed to loose its emission
lines completely.
Therefore, the following results of model atmosphere analysis give only
an approximate estimate of the effective temperature and gravity.

We calculated a small grid of NLTE plane-parallel hydrogen-helium model
atmospheres for all combinations of three values of {\Teff}
($13\,000\K, 14\,000\K, 15\,000\K$) and $\log g$ ($3.0, 3.5, 4.0$)
using a computer code developed by one of us (see, e.g., Kub\'at
\cite{ATAref}, and references therein).
The hydrogen model atom consists of 15 levels of \ion{H}{i} and 1 level
of \ion{H}{ii}, the helium model atom consists of 29 levels of
\ion{He}{i} and 1 level of \ion{He}{ii}.
The atomic data (transition probabilities, photoionization
cross-sections, collisional strengths) were the same as described in
Kub\'at (\cite{spagb}).
By simple comparison of the observed spectrum of higher Balmer lines to
the model one, one finds that the model atmosphere with
$\Teff=14\,000\K$ and $\log g =3.5$ leads to quite a satisfactory
agreement with the observed spectrum for a rotational velocity of
$v \sin i = 170\kms$.

Rotation of the model spectrum was taken into account in an approximate
way using the convolution of the rotation profile with the line profile
following Gray (\cite{gray}).

One can admit that the comparison of the observed and model spectrum
should also take into account the radiation of the secondary component
of the binary.
However, Juza et al. (\cite{Juza91}) presented rather convincing
arguments that the secondary to {\kdra} is not a Roche-lobe filling
giant but a small compact object with a mass probably smaller than some
$0.9~\ms$.
In the whole optical spectrum, the luminosity of such an object would be
negligible in comparison to the Be giant and we conclude that one can
safely neglect the contribution of the secondary to the observed
spectrum in the investigated range of wavelengths.
The values of {\Teff} and $\log g$ we found are not too different from
those presented by Chauville et al. (\cite{tefflogg}) in their Table 1,
namely $\Teff=13900\K$ and $\log g =3.84$, based on the BCD
classification system (Chalonge \& Divan \cite{ChDi52}).

The value of $v \sin i$ is in a good agreement with the value 180{\kms}
determined by Stoeckley \& Buscombe (\cite{rotvel}).
Higher values of the projection rotational velocity seem to be
unrealistic.

The observed and theoretical spectrum is compared in Figures
\ref{HERblue} and \ref{HERred}, which correspond to wavelength regions
covered by the blue and red channel of the {\HEROS} spectrograph,
respectively.
Both parts of the spectrum were obtained simultaneously at
HJD\,51924.5409.

One can see that there is a general agreement between the
\clearpage \noindent
observed and model spectrum in the wings of Balmer lines (with the
exception of {\Halpha}), while the Balmer line cores are affected by the
emission.
Quite satisfactory agreement was also achieved for the \ion{He}{i} lines
and for the \ion{Mg}{ii} 4481\,{\AA} line.
Apart from {\Halpha} wings, which are subject to high variability, the
ratio between \ion{Mg}{ii} 4481\,{\AA} and \ion{He}{i} 4471\,{\AA} is
one of the best classification criteria for spectral types B7 -- B8
(Steele et al. \cite{Steel99}).
However, the synthetic \ion{Si}{ii} 6347 and 6371\,{\AA} lines are
deeper than the observed ones which probably indicates that they are
filled by a weak emission.
A pronounced difference is seen for the infrared \ion{O}{i} 7772, 7774,
7775 triplet, which is observed as a double-peak emission.

The \ion{Fe}{ii} lines studied by Hanuschik (\cite{iron2}) -- namely the
triplet 4924\,{\AA}, 5018\,{\AA}, 5169\,{\AA}, and somewhat weaker
4549\,{\AA}, 4584\,{\AA}, and 5316\,{\AA} lines -- are also observed to
have emission.
On the other hand, there is no observable \ion{Fe}{ii} 6384\,{\AA} line
in the spectra of {\kdra}, similarly as for \object{$\alpha$~Col} from
Hanuschik's sample.

A casual inspection of the observed spectrum could lead to a conclusion
that \ion{Fe}{ii} 4924\,{\AA} line has a P~Cygni profile.
Note, however, that this feature is in fact a blend of \ion{He}{i}
4922\,{\AA} absorption and \ion{Fe}{ii} 4924\,{\AA} double peak
emission.
Similarly, the observed asymmetry of the emission peaks of
\ion{Fe}{ii} 5018\,{\AA} line is also caused by blending with an
absorption \ion{He}{i} line at 5016\,{\AA}.

To see if there are any secular changes in the seemingly photospheric
parts of the observed spectral lines we also compared the theoretical
spectrum with the most recent spectrum of {\kdra} corresponding to
another phase of long-term changes.
It turned out that the only changes are seen in the cores of the
emission lines, especially of lower Balmer lines, and \ion{Fe}{ii}
lines.
The fit to the outer wings of higher lines of the Balmer series remains
unchanged.
This seems to indicate that any parts of the emission-line envelope are
not contributing significantly to the observed continuum radiation or
that their contribution was the same on both occasions.

\section{Measurements}\label{mereni}

To investigate long-term variations of {\kdra}, we focused on {\Halpha},
{\Hbeta}, {\Hgamma}, {\Hdelta}, \ion{Si}{ii} 6347\,\AA,
\ion{Si}{ii} 6371\,\AA, \ion{He}{i} 6678\,\AA,
and \ion{Fe}{ii} 6456\,{\AA} lines.
We measured equivalent width $W$, peak intensities of the double
emission lines, and the central intensity of the absorption reversals
relative to the adjacent continuum ($I_V$, $I_R$ and $I_c$,
respectively) and also the peak separation between violet and red peaks
(\Delnup).
Note that $W$ is taken with a negative sign for the emission line
profile.

Both $W$ and all intensities were measured in the original spectrum,
i.e. without any subtraction of a synthetic spectrum (as it is sometimes
done in similar studies) since this would imply an implicit assumption
of an optically thin envelope.
This need not be the case, especially in the cores of Balmer lines.

Only $W$ and $I_c$ were measured, of course, for the absorption lines of
\element{He} and \element{Si}.
All measurements are collected together with heliocentric Julian dates
of mid-exposures (HJDs) in Tables \ref{Hatable}, \ref{Balpar}, and
\ref{hefesi}.

\subsection{Balmer lines}

\begin{figure}
\centering
\resizebox{\hsize}{!}{\includegraphics{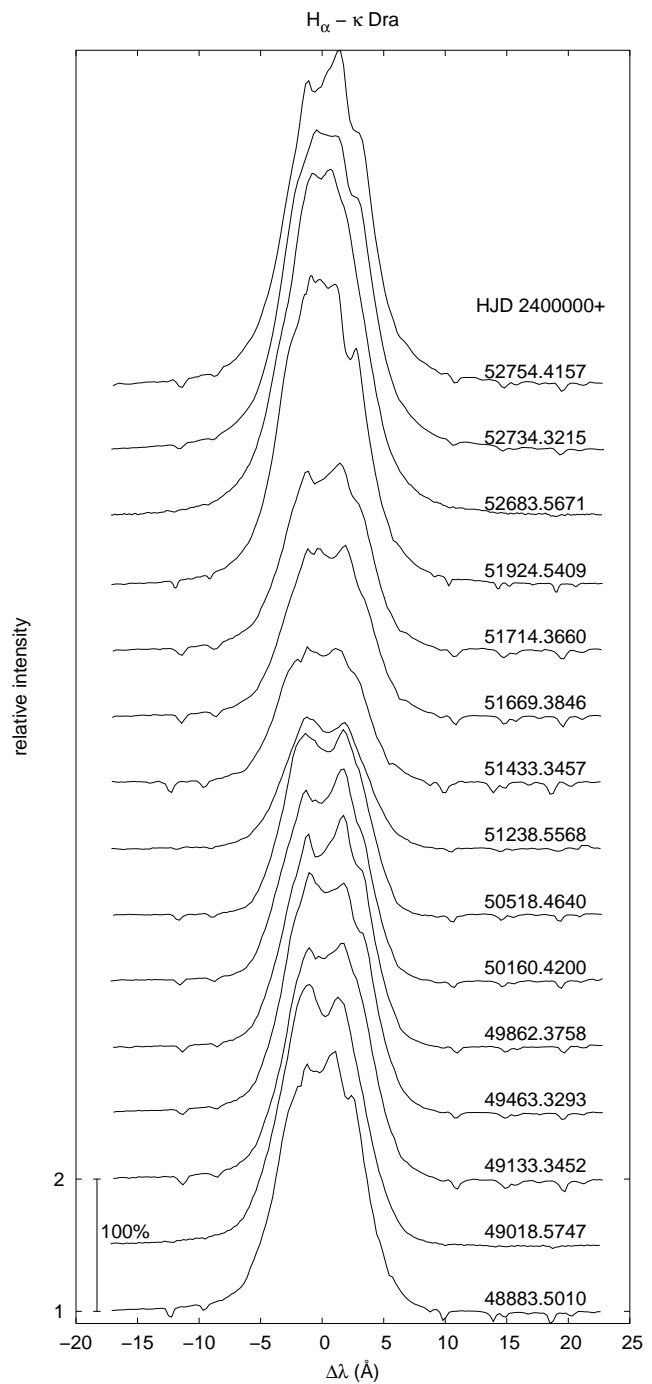}}
\caption{Evolution of the {\Halpha} intensity profile observed in
{\kdra} over the last 11 years.
All spectra are normalized to the local continuum.}
\label{Haprof}
\end{figure}

\begin{figure}
\resizebox{\hsize}{!}{\includegraphics{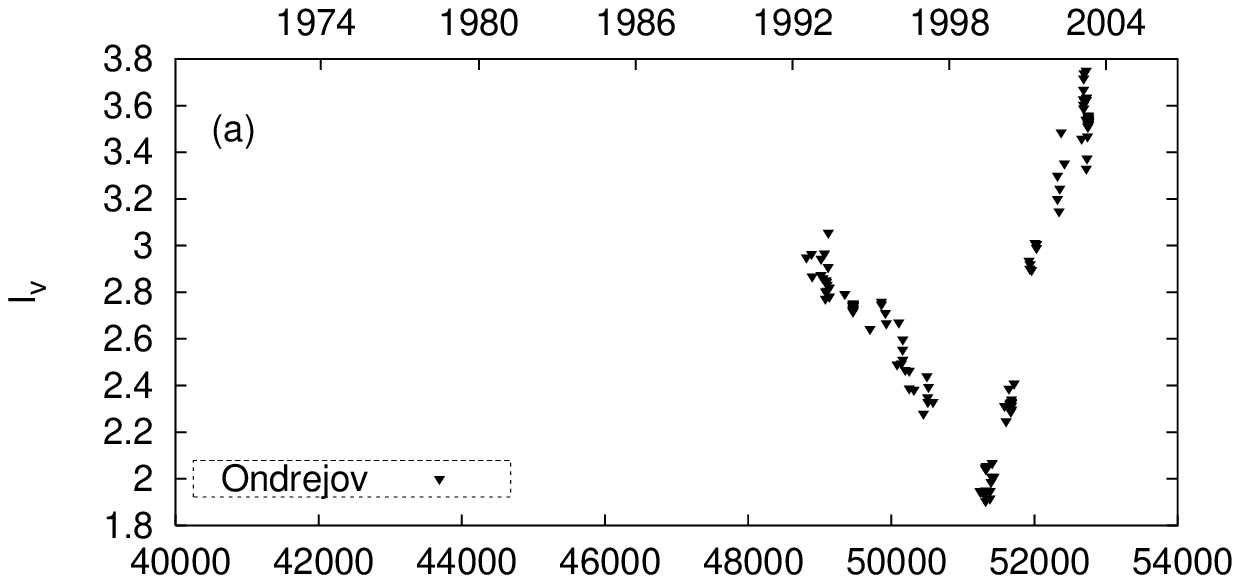}}
\resizebox{\hsize}{!}{\includegraphics{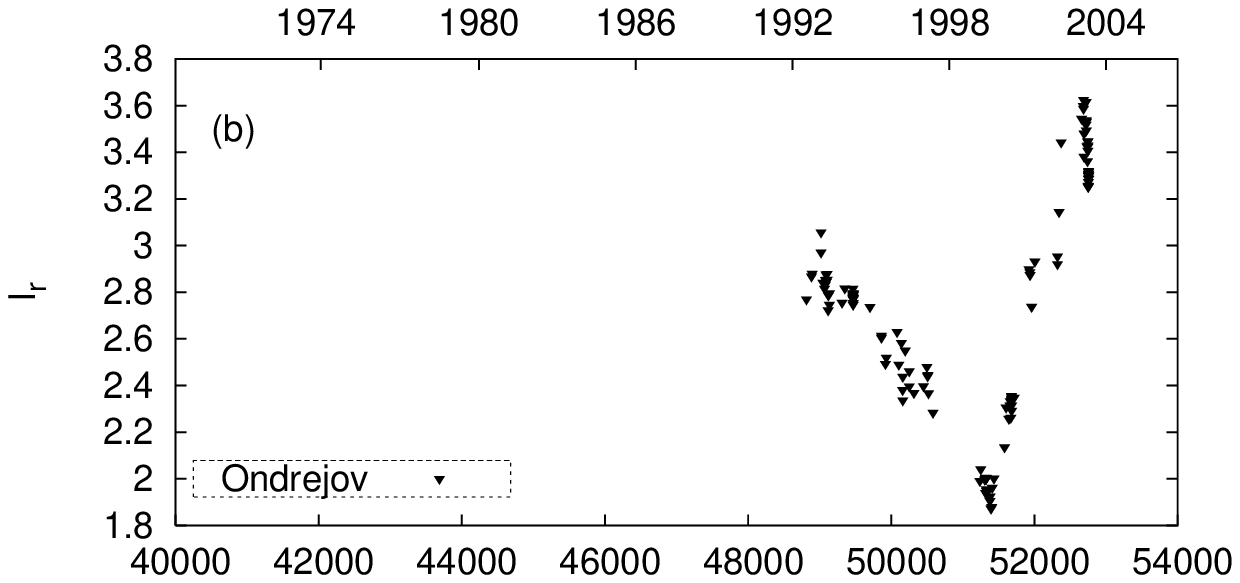}}
\resizebox{\hsize}{!}{\includegraphics{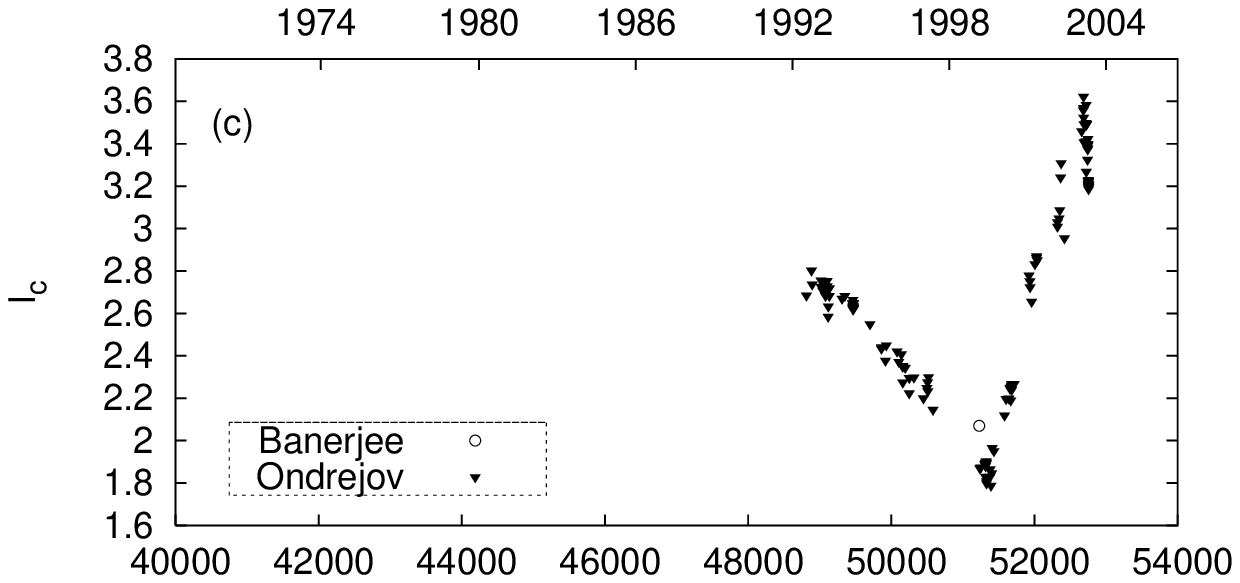}}
\resizebox{\hsize}{!}{\includegraphics{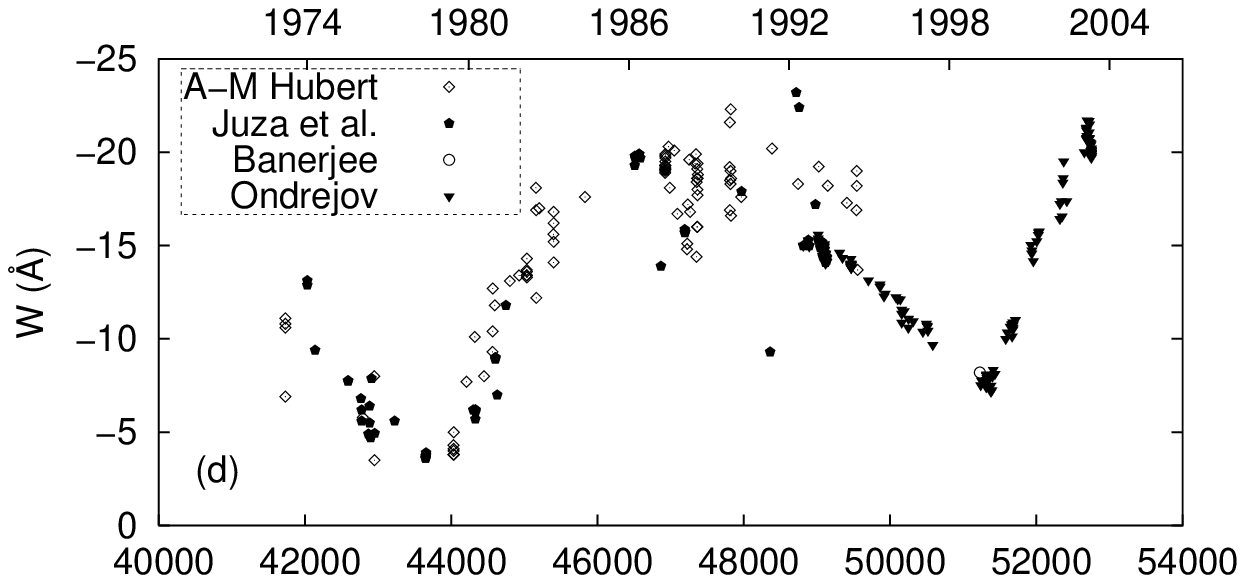}}
\resizebox{\hsize}{!}{\includegraphics{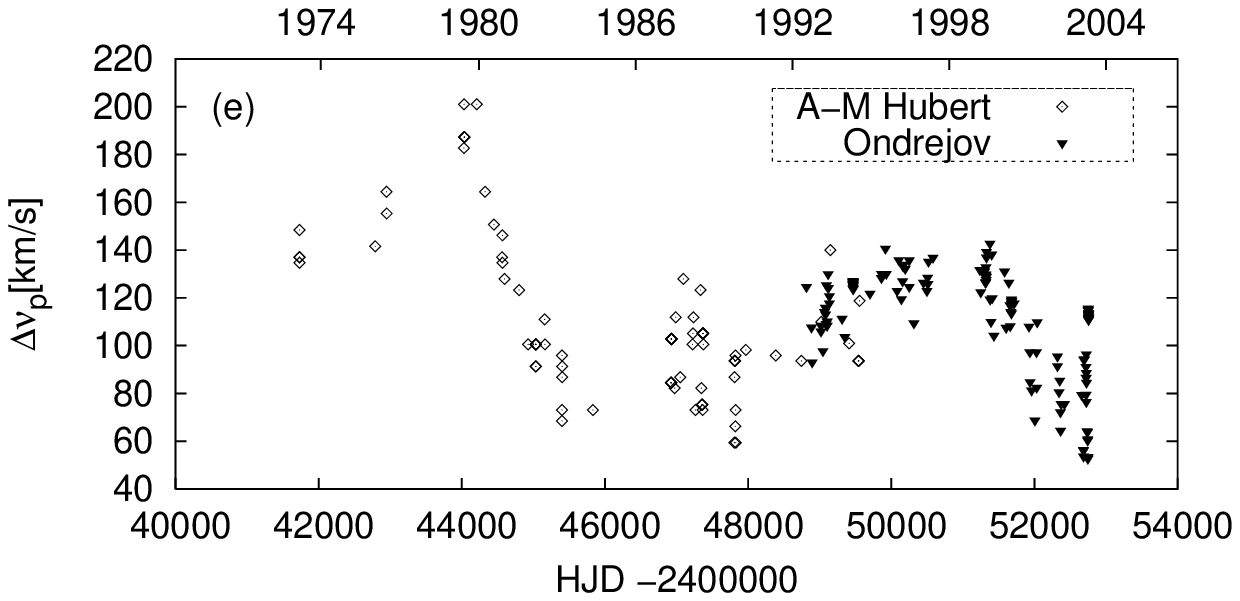}}
\caption{
Correlation between long-term variations of the {\Halpha} emission line.
{\bf \em (a)} and {\bf \em (b):} The relative intensity of the violet
($I_V$) and red ($I_R$) peaks,
{\bf \em (c):} relative central intensity of the line ($I_c$),
{\bf \em (d):} equivalent width of the {\Halpha} line,
{\bf \em (e):} peak-separation \Delnup(\Halpha).}
\label{longvar}
\end{figure}


\begin{figure}
\resizebox{\hsize}{!}{\includegraphics{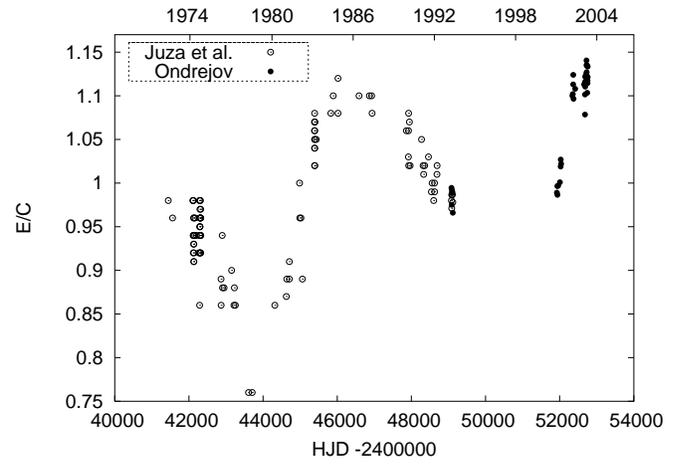}}
\caption{Long-term variations of the $E/C$ {\Hbeta} emission strength
based on both, data from Juza et al. and new observations.}
\label{longphase}
\end{figure}


\begin{figure}
\resizebox{\hsize}{!}{\includegraphics{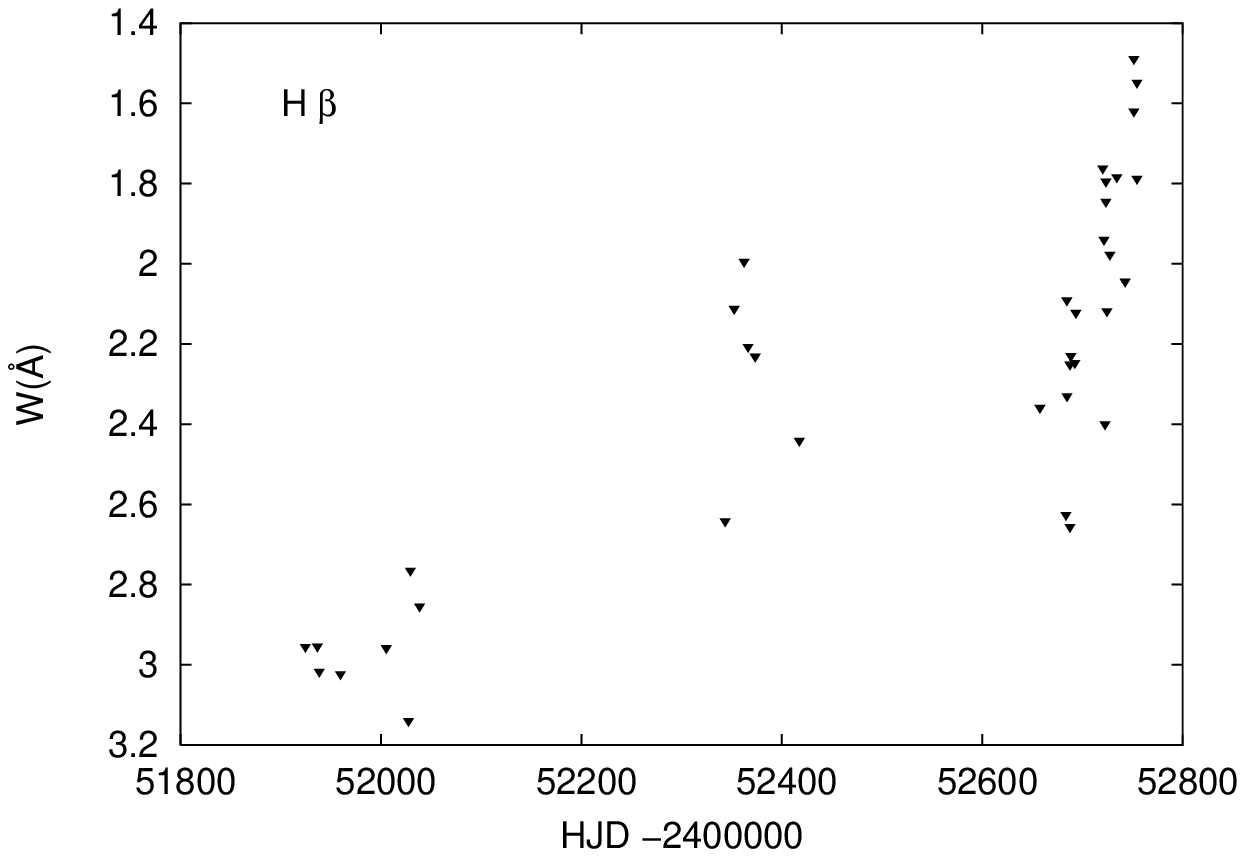}}
\resizebox{\hsize}{!}{\includegraphics{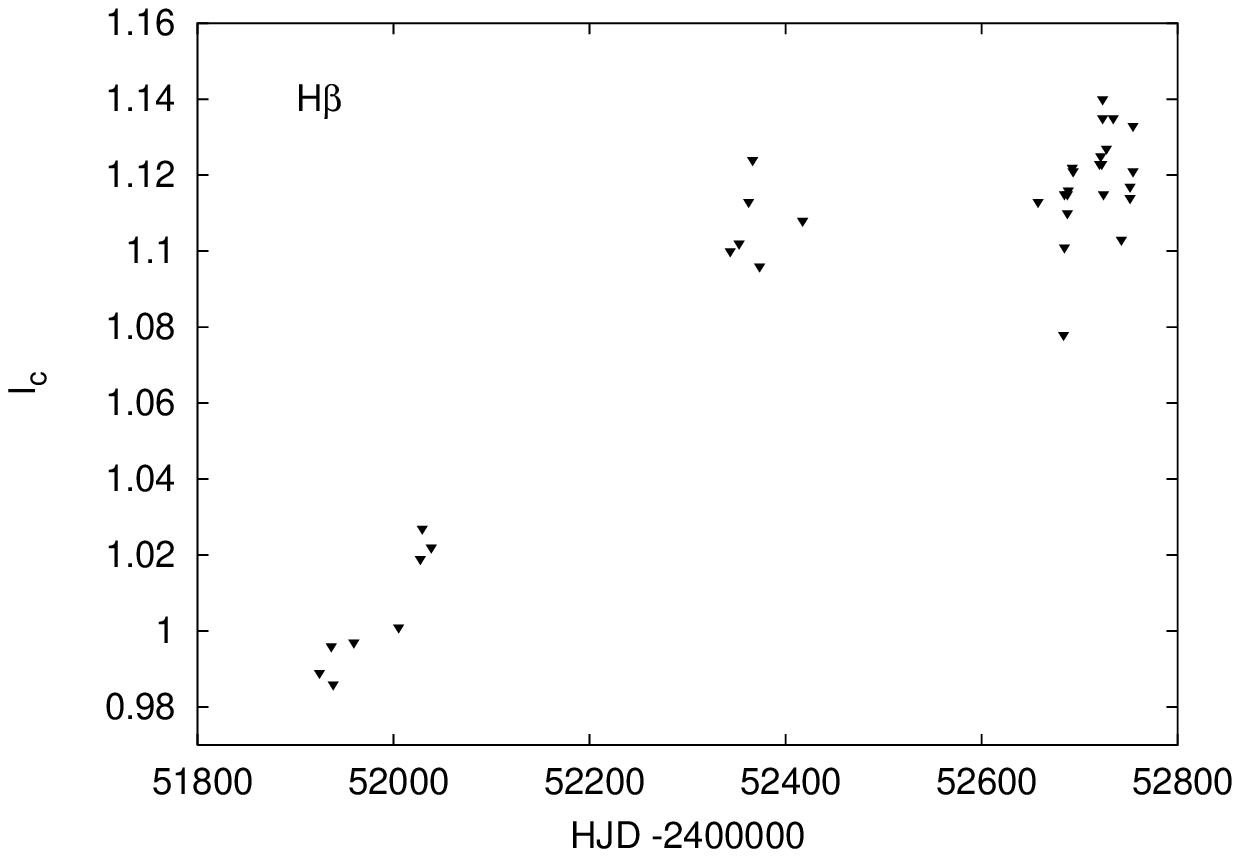}}
\caption{ Variations of {\Hbeta} equivalent width (upper panel) and
intensity (lower panel).}
\label{hbchan}
\end{figure}

\subsubsection{Equivalent width ($W$)}

{\Halpha} line has always been observed as a double-peaked emission with
a variable shape and strength (see Fig.\,\ref{Haprof}).
Fig.\,\ref{longvar}d shows the time variations of the {\Halpha}
equivalent width $W$ during the past 30 years.
Data prior to HJD\,2448802.43819 were taken from Alvarez et al.
(\cite{france90}), Juza et al. (\cite{Juza94}), and Moujtahid et al.
(\cite{AMHdata}).
Numerical values of $W$, shown only in a graphical form in Fig.\,1 of
Moujtahid et al. (\cite{AMHdata}) paper were kindly provided to us by
Dr. A.-M. Hubert.
One measurement was taken from Banerjee et al. (\cite{B00}).
We recall again that the more negative values of the equivalent width
the stronger the emission line.

The observations published earlier locate one minimum of the emission
strength (the arithmetic maximum of $W$) into the interval between
HJD\,2443200 and 2443800, when $\WHalpha\sim-3\mbox{\AA}$.
The other, recent minimum is sharply defined by our data at
HJD\,2451380, when {\WHalpha} reached the value of $-7.17\,\mbox{\AA}$.
With some scatter, the available data locate the time of the maximum
emission around HJD\,2447000 with a value of {\WHalpha} between
$-20\,\mbox{\AA}$ and $-21\,\mbox{\AA}$.
The most recent data again indicate values near
$\WHalpha=-21\,\mbox{\AA}$ or lower.

Available equivalent widths of a {\Hbeta} lines basically confirm the
trend defined by {\Halpha}, though with an increased scatter.
This is illustrated by the upper panel of a Fig.\,\ref{hbchan}.

\subsubsection{Central ($I_c$) and peak ($E/C$) line intensities}

\begin{figure}
\resizebox{\hsize}{!}{\includegraphics{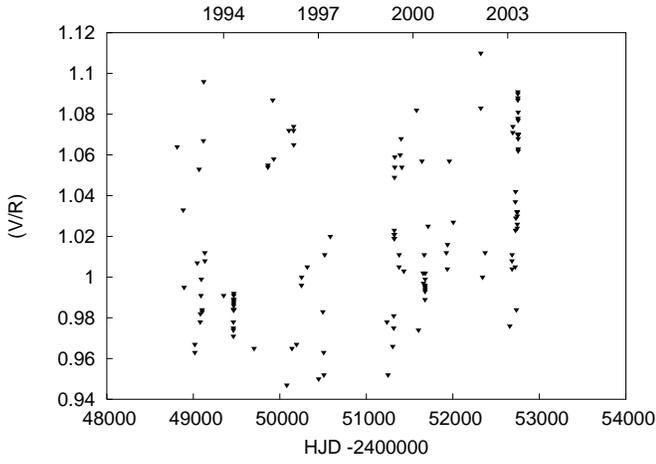}}
\caption{A time plot of the $V/R$ changes of {\Halpha} emission line
during the last 11 years. No long-term variations are observed.}
\label{vrlong}
\end{figure}

The central intensity $I_c$ of the emission line we define as the
intensity of the central part of the line relative to the continuum
level.
For the particular case of our star (\kdra) we measure it at the
position of the central absorption.
On the other hand, we can express the peak intensity by the ratio $E/C$,
which is the mean intensity of both red and violet peaks, i.e.
$(I_V+I_R)/2$.
We measured the central intensities $I_c$ for each line.
For comparison with the data of Juza et al. (\cite{Juza94}) we measured
the intensity of {\Hbeta} in terms of $(I_V+I_R)/2$ as well.
Our measurements of the central intensity of {\Halpha} were
complemented by one observation published by Banerjee et al.
(\cite{B00}).
The central intensity $I_c$ shows parallel behavior with red ($I_R$) and
violet ($I_V$) peak intensities, and also with {\WHalpha}, as
illustrated by the first four panels of Fig.~\ref{longvar}.

Fig.\,\ref{longphase} represents the intensity variation of {\Hbeta}
emission over the last 30 years.
Data prior to HJD\,2451924.5409 have been adapted from Juza et al.
(\cite{Juza94}).
One can see a continuing cyclic variation over which a secular increase
in the emission strength seems to be overlapped.

The central intensities of {\Hgamma} and {\Hdelta} lines have been
slowly increasing with time from 0.62 to 0.66, and from 0.53 to 0.56,
respectively, over the period covered by our new data, showing again
changes parallel to those seen in {\Halpha}.

Parallel changes of the intensity of the emission peaks $I_V$ and $I_R$
have been found for {\Halpha}.
Note, however, that -- in contrast to many other Be stars -- there is no
clear evidence of parallel long-term $V/R$ variations as
Fig.~\ref{vrlong} shows.

As shown already by Juza et al. (\cite{Juza91}), this ratio exhibits
phase-locked variations with phase of the 61\fd55 binary orbit.
We postpone investigation of orbital changes of {\kdra} into a follow-up
study.

\subsubsection {Peak separation (\Delnup)}

\begin{figure}
\resizebox{\hsize}{!}{\includegraphics{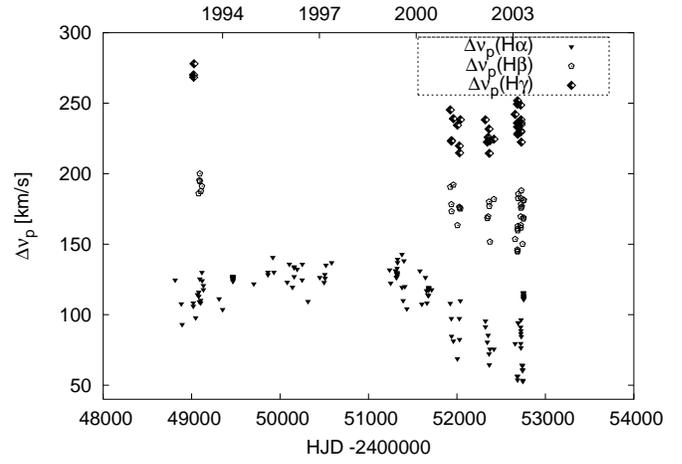}}
\caption{Comparison of the peak separation of {\Halpha},
{\Hbeta}, and {\Hgamma} lines.}
\label{balall}
\end{figure}

Usually the double-peak structure is one of the characteristic features
of the emission lines in Be stars.
Peak-separations (\Delnup) were measured for Balmer lines whenever they
were available in our {\kdra} spectra, the most prominent ones were in
{\Halpha}, {\Hbeta}, and {\Hgamma} lines.

The {\Delnup} of {\Halpha} underwent very clear long-term variations
over the interval covered by our data.
After HJD\,48802.43819 (in 1992) it started a gradual increase until it
reached a maximum of 142{\kms} at HJD~51378.5441 (in 1999).
This was followed by a steep decrease to its minimum value of 53{\kms}
on HJD\,52684.5418 (in 2003).
Note that Arsenijevi\'c et al (\cite{france94}) reported the maximum
value of the peak separation of 206{\kms} (in the years 1961--1962 and
1985--1986) and the minimum of 69{\kms} in 1980.
Fig.\,\ref{longvar}e displays the long-term behaviour of {\Halpha} peak
separation based on the new and compiled data.
The first part of the {\Delnup} long-term behaviour is mainly based on
data from Moujtahid et al. (\cite{AMHdata}), kindly provided to us by
Dr.~A.-M.~Hubert.
They go from early 70's to 1995.

We also measured the peak separation for {\Hbeta} and {\Hgamma} lines in
all spectrograms, where these lines were available.
Fig.~\ref{balall} illustrates the comparison between the peak separation
values of {\Halpha}, {\Hbeta} and {\Hgamma}.
The measurements of $\Delnup(\Hgamma)$ have large uncertainties because
of weakness of the emission.
Note that the values of $\Delnup(\Hbeta)$ are significantly higher than
those of {\Halpha}, and smaller than $\Delnup(\Hgamma)$.
In particular, $\Delnup(\Hgamma)$ ranges from 220 to 250{\kms},
$\Delnup(\Hbeta)$ from 144 to 190{\kms} and $\Delnup(\Halpha)$ from 53
to 108{\kms} for the time interval where data for all three lines
overlap.
We observe that at any time the relation
\begin{equation}
\Delnup(\Halpha) < \Delnup(\Hbeta) < \Delnup(\Hgamma)
\end{equation}
\noindent holds.
This reflects different regions of the formation of individual Balmer
emission lines.
The opacity in lower members of Balmer series is larger.
Consequently, the lines are optically thick in the radial direction to a
relatively large distance.
This distance decreases with the increasing principal quantum number of
the upper transition level.
Higher members of the series originate at regions closer to the stellar
surface, having higher Keplerian velocity, and therefore larger
{\Delnup}.

Averaging the ratios $\Delnup(\Hgamma)/\Delnup(\Halpha)$,
$\Delnup(\Hgamma)/\Delnup(\Hbeta)$, and
$\Delnup(\Hbeta)/\Delnup(\Halpha)$ over all common epochs we can
conclude that
\begin{subequations}
\begin{equation}
\frac{\Delnup(\Hgamma)}{\Delnup(\Halpha)} = 1.37\pm0.15
\end{equation}
\begin{equation}
\frac{\Delnup(\Hgamma)}{\Delnup(\Hbeta)} = 2.98\pm0.76
\end{equation}
\begin{equation}
\frac{\Delnup(\Hbeta)}{\Delnup(\Halpha)} = 2.16\pm0.32
\end{equation}
\end{subequations}
Within the errors, these ratios are close to the mean values estimated
by Hanuschik et al. (\cite{H88}) and Dachs et al. (\cite{D92}):
\begin{subequations}
\begin{equation}
\frac{\Delnup(\Hbeta)}{\Delnup(\Halpha)} \approx 1.8\pm0.5
\end{equation}
\begin{equation}
\Delnup(\Hgamma) \approx 1.2\Delnup(\Hbeta) \approx 2.2\Delnup(\Halpha),
\end{equation}
\end{subequations}
for Be stars having symmetric double-peak profiles.

\subsection{Other lines}

\begin{figure}
\centering
\resizebox{\hsize}{!}{\includegraphics{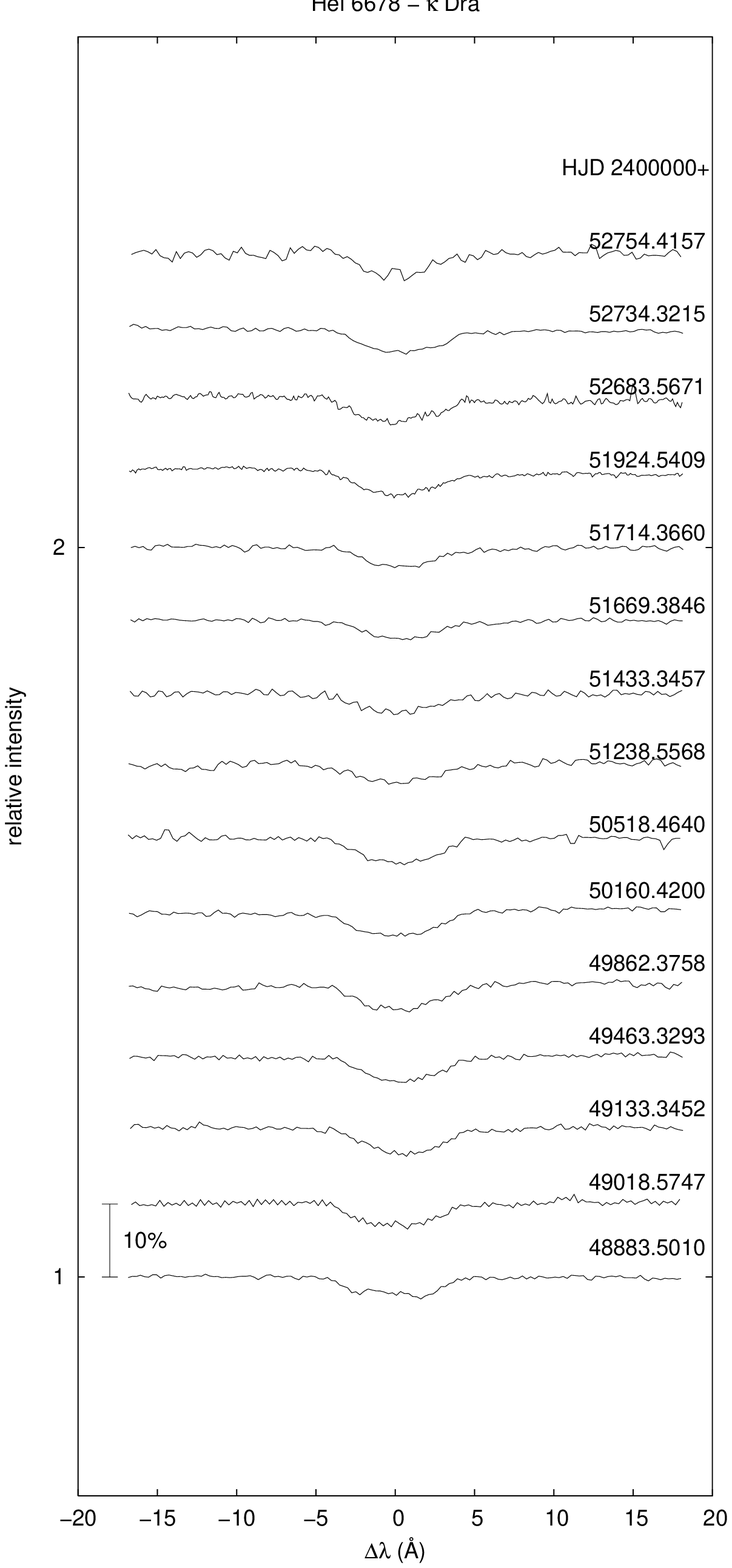}}
\caption{Evolution of the \ion{He}{i} 6678\,{\AA} intensity profile
observed in {\kdra} over the last 11 years.
All spectra are normalized to the local continuum.}
\label{he1sp}
\end{figure}

\begin{figure}
\resizebox{\hsize}{!}{\includegraphics{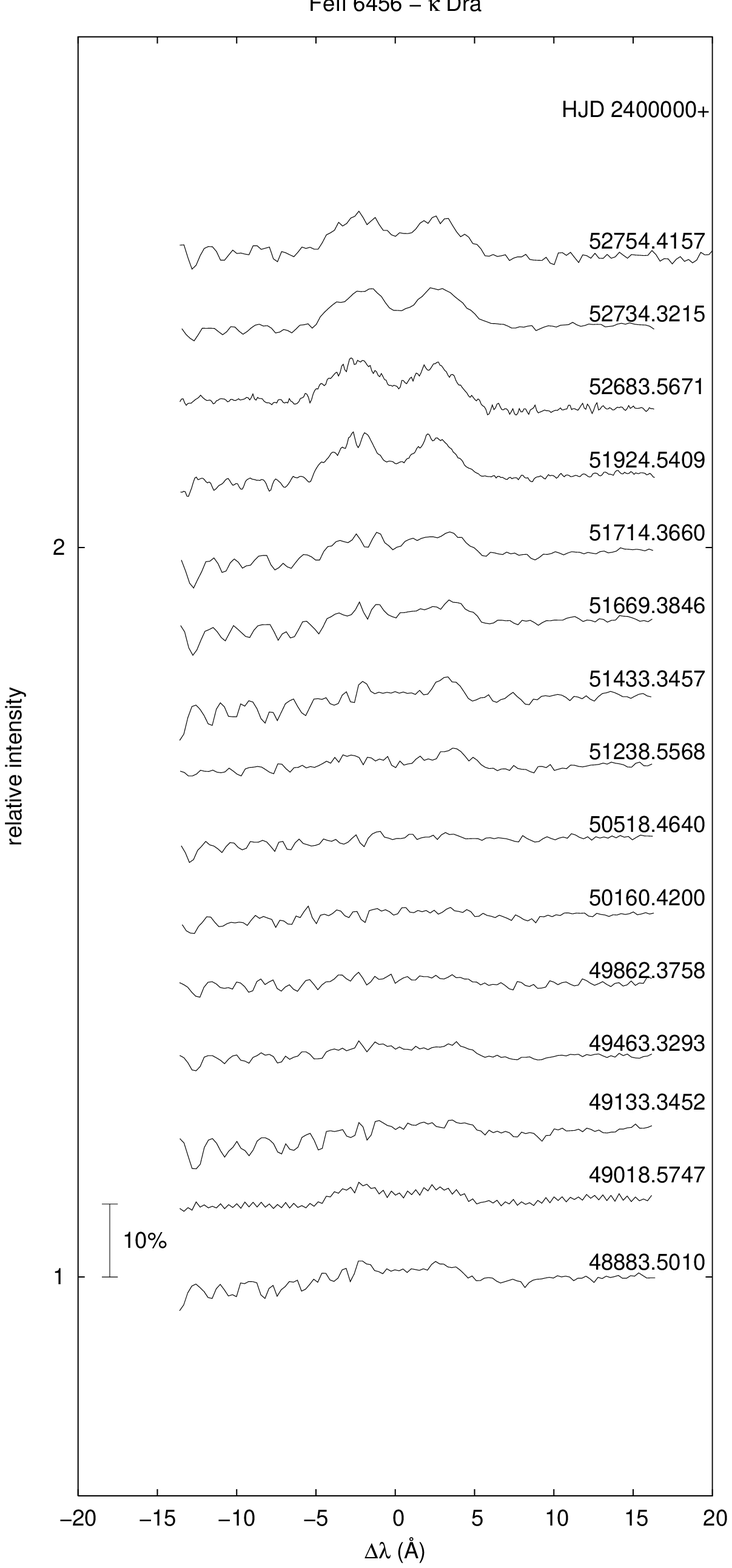}}
\caption{Evolution of the \ion{Fe}{ii} 6456\,{\AA} intensity profile
observed in {\kdra} over the last 11 years.
All spectra are normalized to the local continuum.}
\label{fe2sp}
\end{figure}

\begin{figure}
\resizebox{\hsize}{!}{\includegraphics{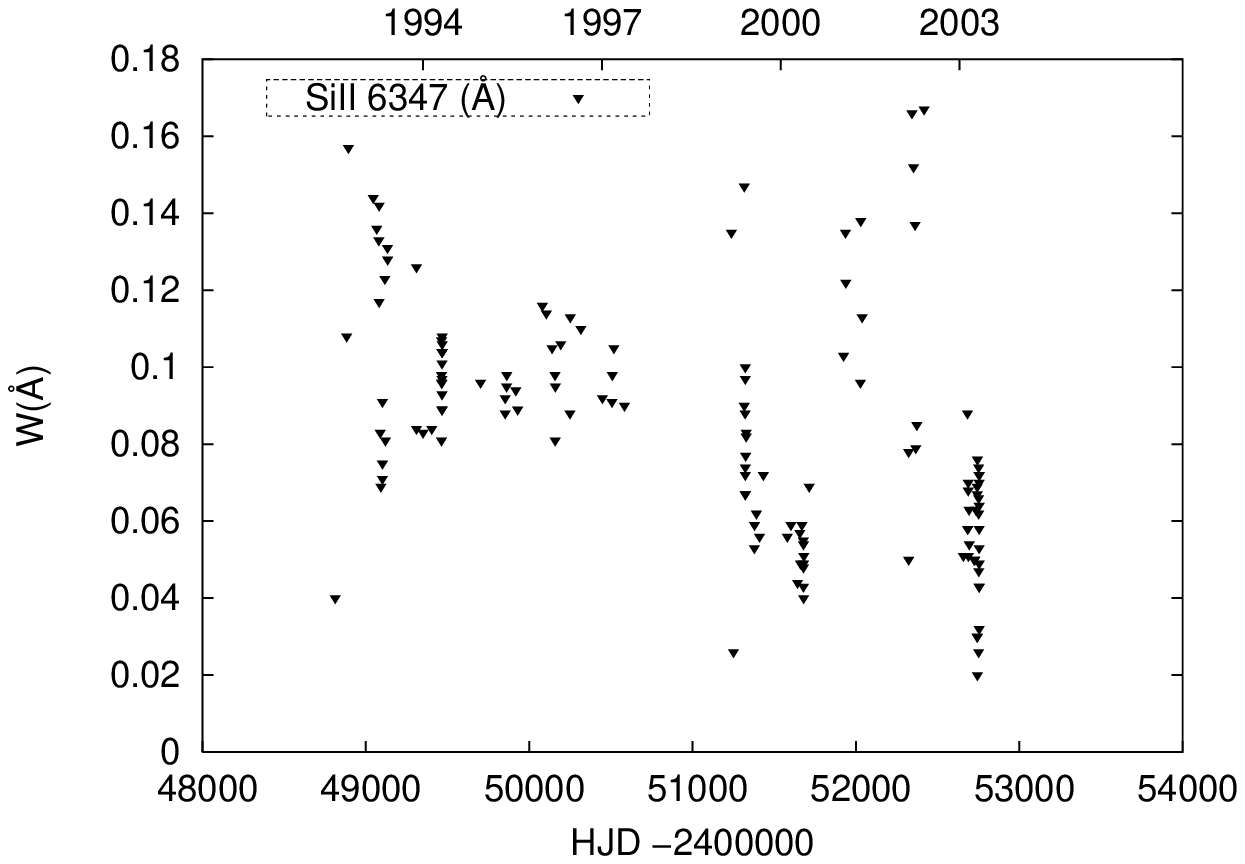}}
\resizebox{\hsize}{!}{\includegraphics{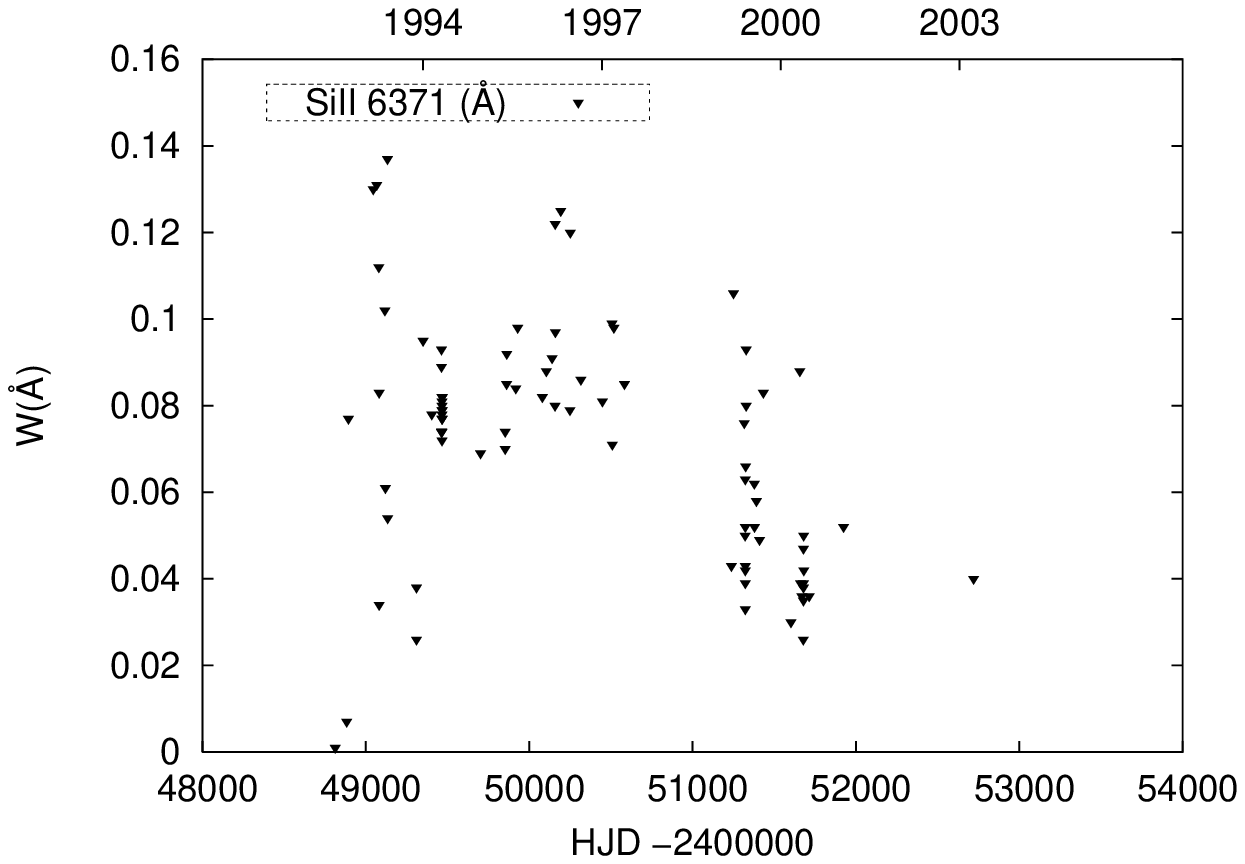}}
\resizebox{\hsize}{!}{\includegraphics{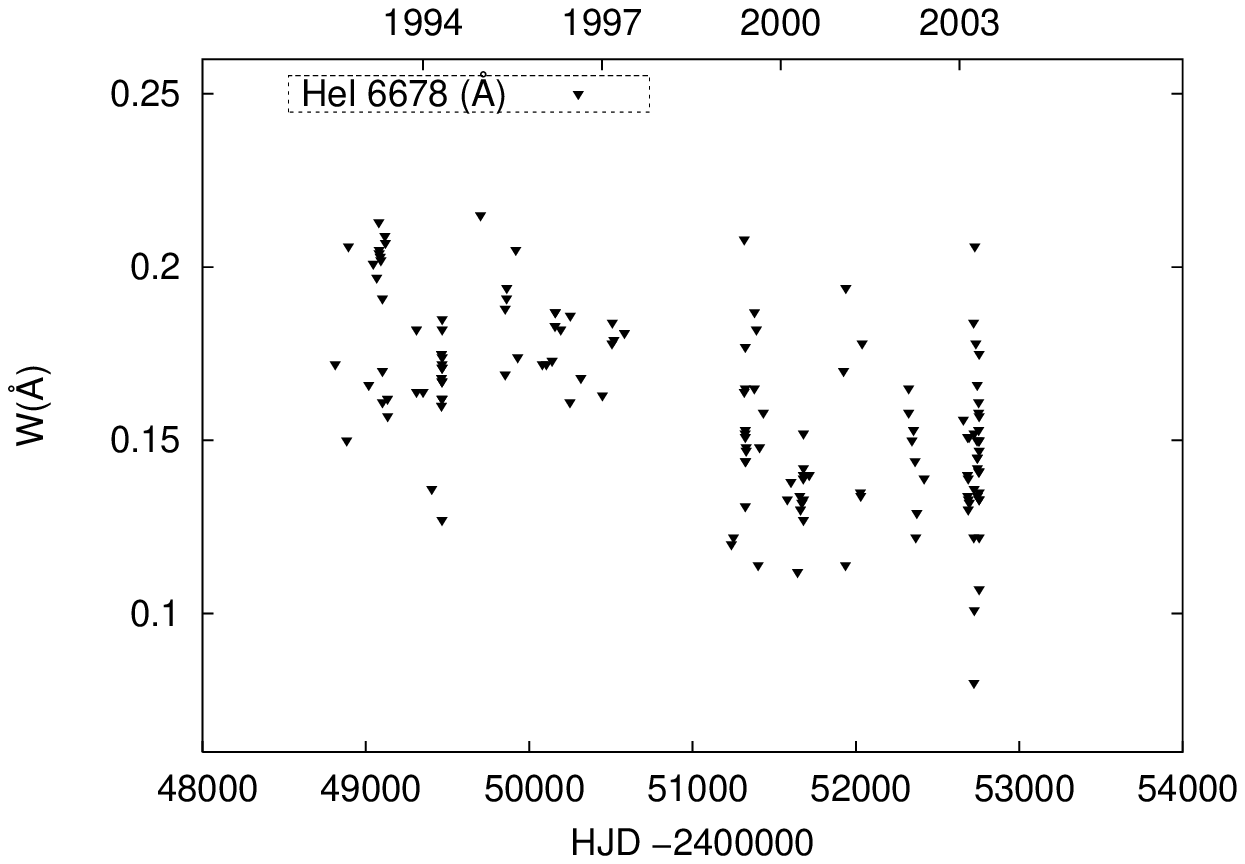}}
\caption{Long-term variations of the equivalent width of \ion{Si}{ii}
6347\,{\AA} line (upper panel), \ion{Si}{ii} 6371\,{\AA} line
(middle panel), and \ion{He}{i} 6678\,{\AA} (lower panel).}
\label{varsi}
\end{figure}

\begin{figure}
\resizebox{\hsize}{!}{\includegraphics{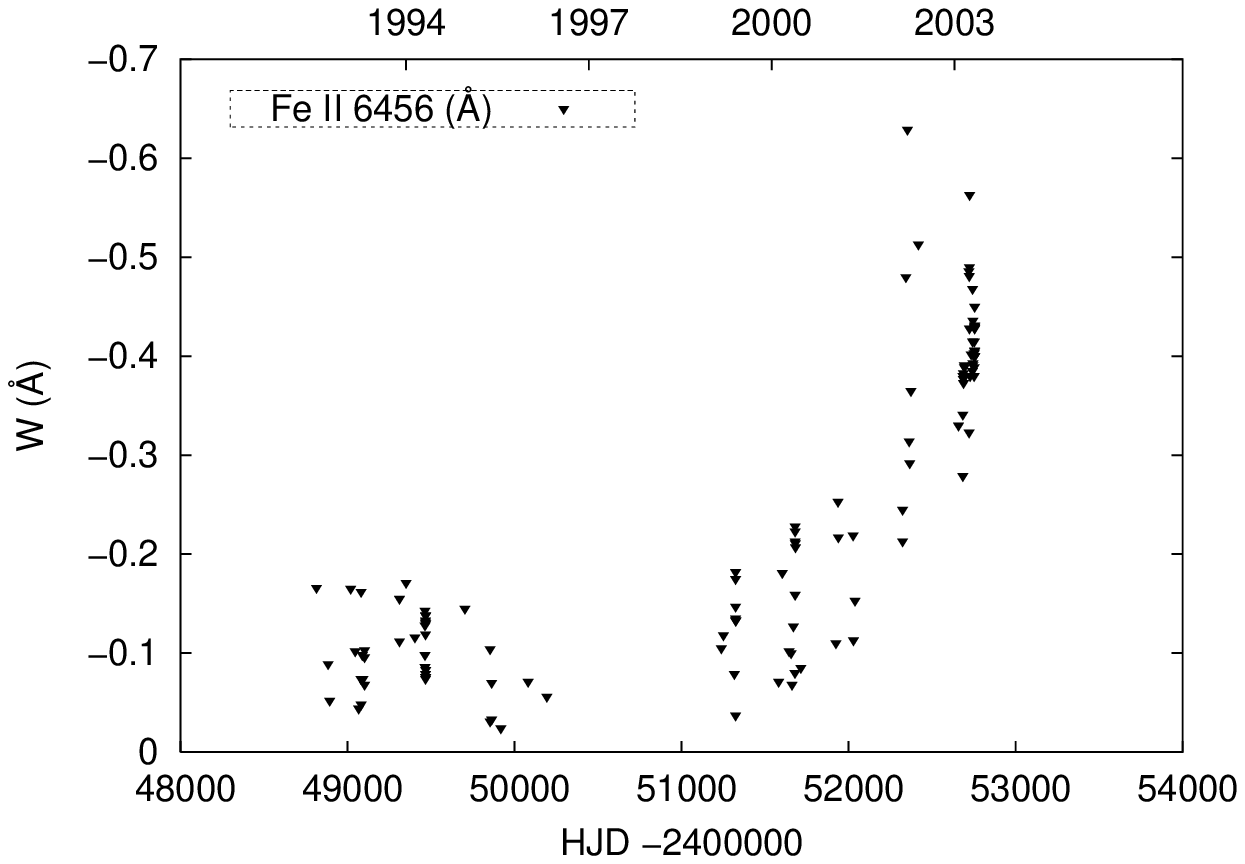}}
\resizebox{\hsize}{!}{\includegraphics{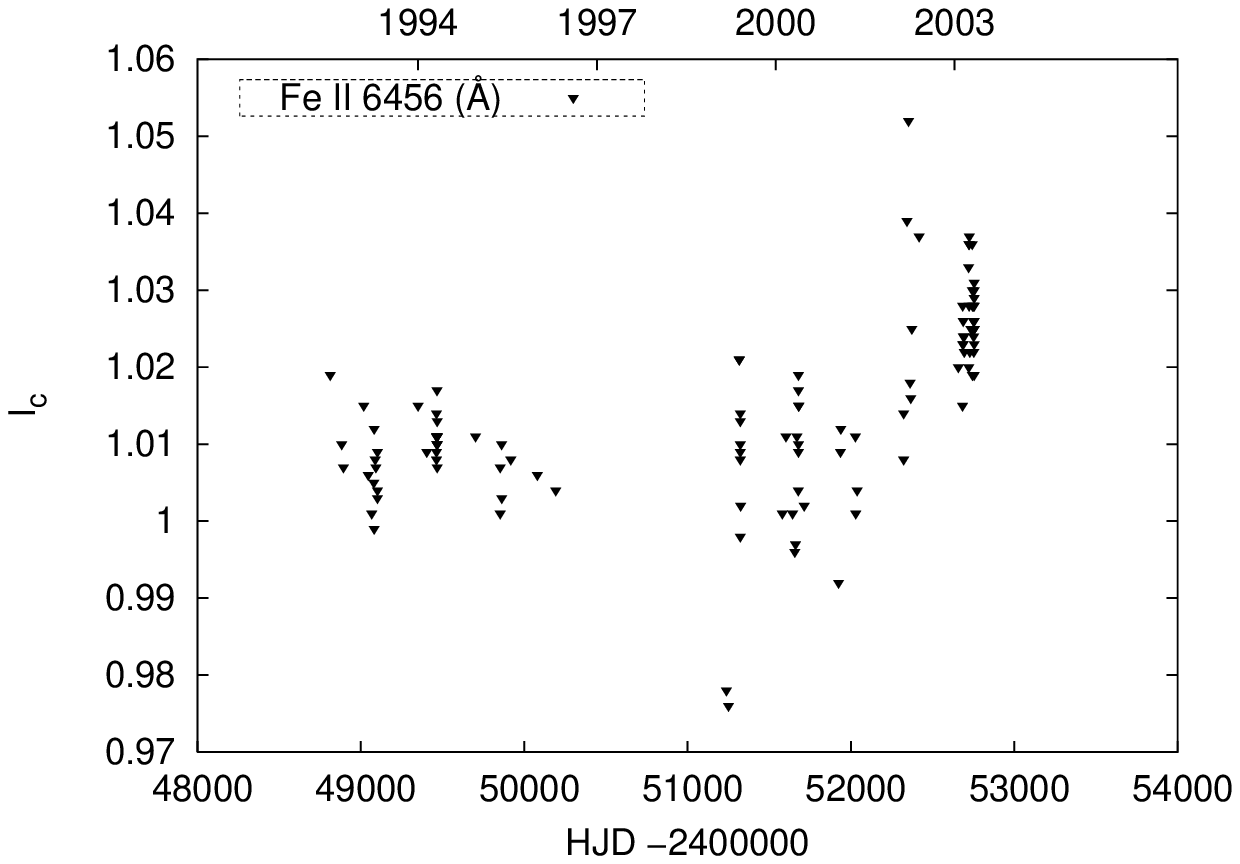}}
\caption{Long-term variations of \ion{Fe}{ii} 6456\,{\AA} line
equivalent width (upper panel) and intensity (lower panel).}
\label{variron}
\end{figure}

The lines of \ion{Si}{ii} 6347\,{\AA}, \ion{Si}{ii} 6371\,{\AA},
\ion{Fe}{ii} 6456\,{\AA}, and \ion{He}{i} 6678\,{\AA} are all relatively
weak in comparison to the Balmer lines but in most cases their
measurements were reliable enough.
The silicon lines \ion{Si}{ii} 6347\,{\AA} and \ion{Si}{ii} 6371\,{\AA},
and the helium line \ion{He}{i} 6678\,{\AA} have been in absorption,
while the \ion{Fe}{ii} 6456\,{\AA} line has appeared as a double-peaked
emission profile during all 11~years of observations.
The line profiles of \ion{He}{i} 6678\,{\AA} and
\ion{Fe}{ii} 6456\,{\AA} are shown in Figs.~\ref{he1sp} and \ref{fe2sp},
respectively.
Note that at certain times it was quite difficult to measure these weak
lines because of a heavy blending with neighbouring telluric lines.

\ion{He}{i} 6678\,{\AA} is the strongest absorption line besides
{\Halpha} in the red part of the spectrum.
It appears as a shallow absorption line with an intensity 0.04 relative
to the continuum level (see Fig.~\ref{he1sp}).
Its equivalent width varies from 0.12\,{\AA} to 0.22\,{\AA} from the
early 1992 to 1999 and then decreases, with large fluctuations, to
0.1\,{\AA} (see lower panel of Fig.~\ref{varsi}).
In spite of the scatter, one can see that \ion{He}{i} 6678\,{\AA}
equivalent width follows the same trend as the emission lines
of {\Halpha} and \ion{Fe}{ii}~6456\,{\AA}.

Both \ion{Si}{ii} 6347\,{\AA} and \ion{Si}{ii} 6371\,{\AA} lines
exhibit variations in their equivalent width from 0.04\,{\AA} to
0.1\,{\AA} at most, as shown in the upper and middle panels of
Fig.~\ref{varsi}.
In spite of a considerable scatter, one can recognize a monotonous
decrease of their equivalent width with time.

Line profile of \ion{Fe}{ii} 6456\,{\AA} appears permanently as a double
emission (see Fig.~\ref{fe2sp}).
Its central intensity $I_c$ is gradually varying between 0.98 and 1.06
and back, in parallel to the variations of the Balmer lines;
The equivalent width of this line increases (with fluctuations)
from the value of $-0.17$\,{\AA} in 1992 to $-0.05$\,{\AA} in 1999
(see Fig.~\ref{variron}).
Between 2000 and 2003 the emission strengthens rapidly so that the
equivalent width arithmetically decreases, reaching its lowest value of
$-0.5$\,{\AA} during the most recent observations.
The epochs when the \ion{Fe}{ii} 6456\,{\AA} line is either very faint
or completely missing, correlates well with phases of weakest {\Halpha}
emission.

Central intensities of \ion{He}{i} 6678\,{\AA} line, and of
both \ion{Si}{ii} 6347\,{\AA} and 6371\,{\AA} lines are subject to a
large scatter and are not displayed.

\subsection{Photometry}

\begin{figure}
\resizebox{\hsize}{!}{\includegraphics{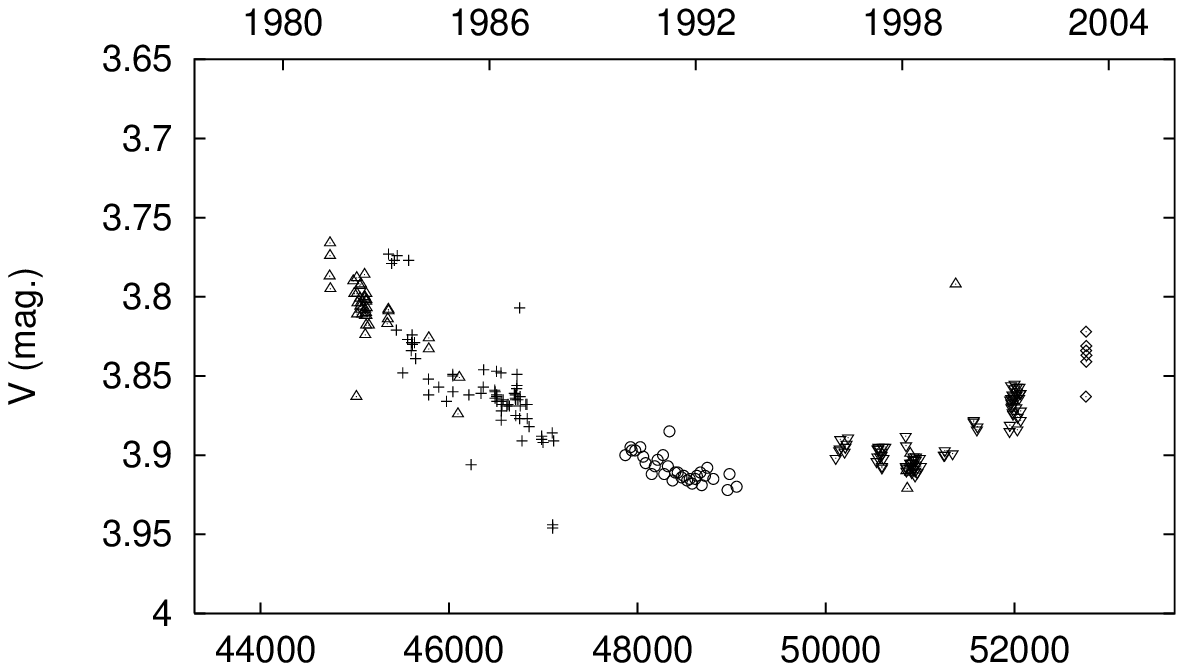}}
\resizebox{\hsize}{!}{\includegraphics{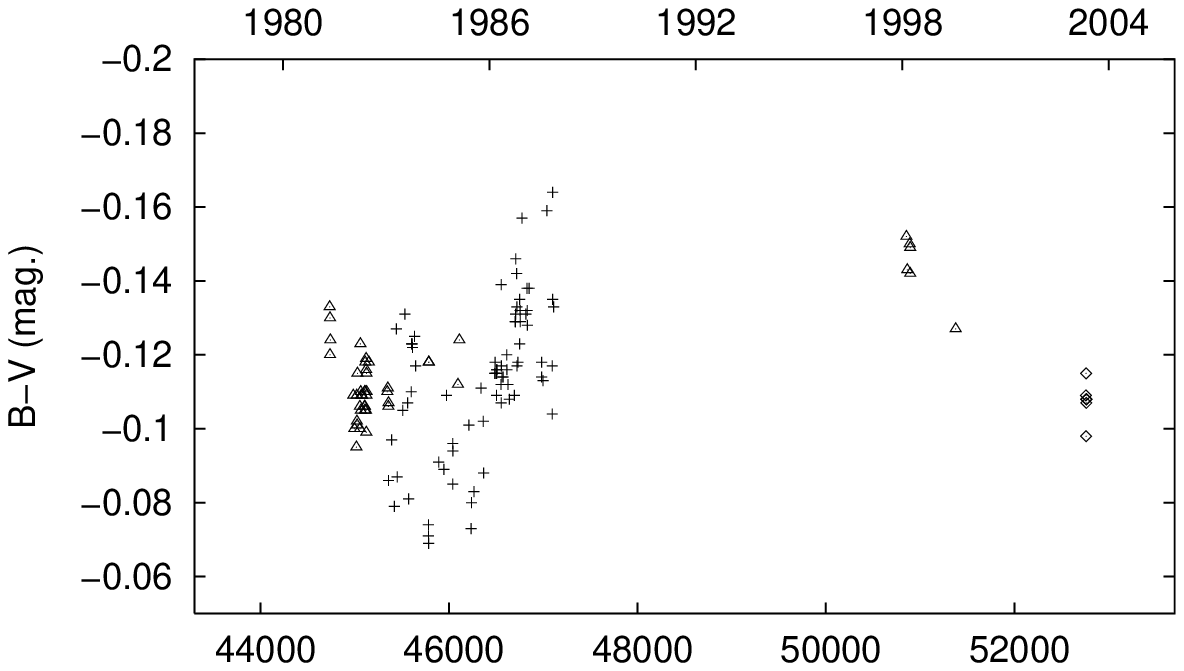}}
\resizebox{\hsize}{!}{\includegraphics{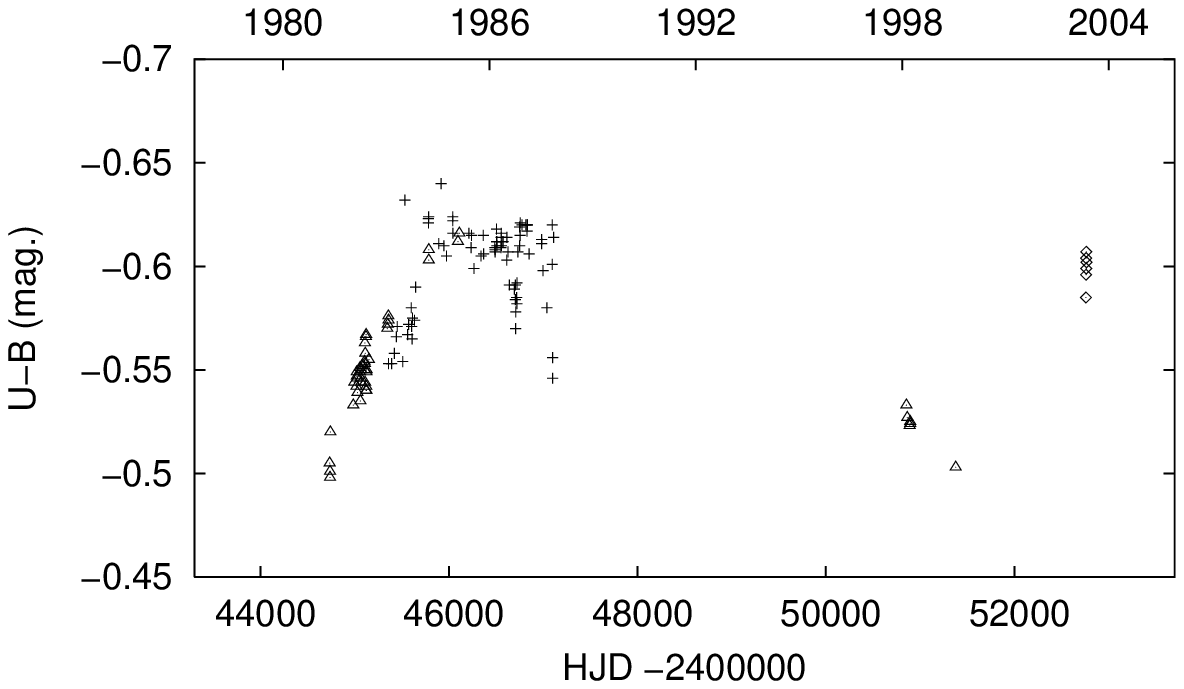}}
\caption{Photometry of {\kdra} from Hvar ($\vartriangle$), Skalnat\'e
Pleso ($+$), and Tubitak ($\diamond$) observatories, from the Hipparcos
satellite ($\circ$) and from Croxall et al. (\cite{karen} --
$\triangledown$).}
\label{photo}
\end{figure}

Long-term variations of the $V$ magnitude and of both {\bv} and {\ub}
indices are shown in the Fig.~\ref{photo} in the form of 1-d normals.
One can see that the brightness in the $V$ passband clearly correlates
with the long-term spectral changes.

Besides, it is clear that occasional smaller brightenings or fadings
occur on a much shorter time scale than the long-term ones.
While it is conceivable that a single case of such a behavior could be
blamed for larger observational errors, we want to point out that, e.g.,
the sudden brightening observed at Hvar on HJD~2451377.4, is undoubtedly
a real phenomenon.
This normal point is based on three observations giving a small rms
error of the mean, obtained on a good night with measured extinction and
normal values observed for the check star.
Such brightenings were also observed for some other long-term Be
variables, e.g. {$\omega$~CMa} (HD\,56139) -- see Fig.\,8 in Harmanec
(\cite{hec98a}).

\section{Basic physical properties of {\kdra}}

\begin{figure}
\resizebox{\hsize}{!}{\includegraphics{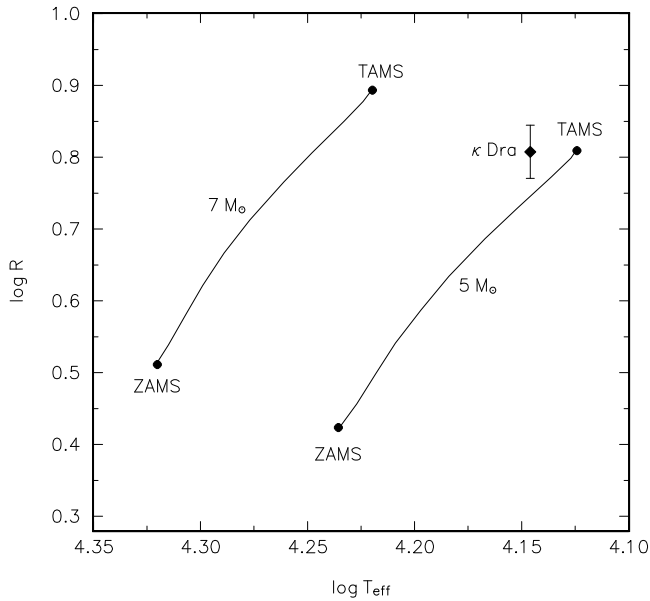}}
\caption{The position of {\kdra} in the log~$R$ vs. log~$T_{\rm eff.}$
diagram. Schaller's et al. (\cite{schaller}) evolutionary models for
5~{\ms} and 7~{\ms} stars are shown by solid lines.}\label{logr}
\end{figure}

Most of published spectral classifications of {\kdra} agree on
the spectral class B5IIIe--B6IIIe -- cf., e.g., Slettebak (\cite{S82}),
Hoffleit \& Jaschek (\cite{HJ82}).
The effective temperature derived here from the line-profile modelling
corresponds to spectral class B6.

One may carry out several consistency checks.
First, one can use the accurate Hipparcos parallax of {\kdra}
$(6.55\pm0.55)\,\mathrm{mas}$ (Perryman et al. \cite{hip}) and a
properly chosen $V$ magnitude to an~estimate of the stellar radius.
As already mentioned, the star has never been observed without emission.
However, in the period with only a very weak emission around
HJD~2443700 the $V$ magnitude of the object attained a minimum
brightness of about 3\fm97.
Interstellar reddening of {\kdra} was derived by Beeckmans \&
Hubert-Delplace (\cite{ebv}).
Three different methods consistently led to $E(B-V)$=0\fm04.
Using the unreddened magnitude $V_0$=3\fm93, bolometric correction from
Code et al.  (\cite{code}), and the Hipparcos parallax, one obtains
\begin{equation}\label{polomer}
R=6.4\pm0.5 {\rm R}_{\sun}.
\end{equation}
The spectroscopic mass which follows from the comparison
with the model atmosphere and $\log g = 3.5$ is then
\begin{equation}\label{hmota}
M=4.8\pm0.8 {\rm M}_{\sun}.
\end{equation}
Errors in the Eqs. \eqref{polomer} and \eqref{hmota} reflect the error
of the parallax.

In Fig.\,\ref{logr} we compare the stellar radius derived above with
Schaller's et al. (\cite{schaller}) evolutionary tracks for 5 and
7~{\ms} models stars.
One can see that the position of {\kdra} corresponds to a giant star,
evolved close to the terminal-age main sequence (TAMS)
and having a mass slightly over 5\,{\ms}.
For solar composition, the corresponding evolutionary age is something
like $8\cdot10^{7}$ years.
The bolometric magnitude of the star is $-3\fm1$.

It is encouraging to see that in spite of all uncertainties, we arrived
at consistent estimates of the mass of the Be star by two
{\em independent} methods.
This is very important since just the lack of direct determinations of
stellar masses of Be stars is one of the sources of uncertainty in
considerations about how close the rotation of Be stars is to break up
velocity at the equator.

\section{Character of long-term variations}

During the past century, {\kdra} has shown very pronounced spectral
variability.
The strength of {\Hbeta} emission was found to vary on a scale of about
two decades.
While such variations of Be stars were usually found to be cyclic with
variable length of individual cycles, in the case of {\kdra} several
authors argued in favour of a real periodicity.
Periods of 30 and 23 years for the variation of $E/C$ of {\Hbeta}
have been suggested by Struve (\cite{St25}) and Jessup (\cite{J32}),
respectively.
McLaughlin (\cite{Mc37}) noted that in 1936 the emission
at {\Hbeta} and {\Halpha} had almost disappeared.
Alvarez et al. (\cite{france90}) reported that the {\Halpha} line
equivalent width varies between 5\,{\AA} and 25\,{\AA}.
As their Figure 1 shows, they also found a clear correlation between
continuum polarization and {\WHalpha}.
This correlation was later confirmed by a large series of observations
published by Arsenijevi\'c et al. (\cite{france94}).

Juza et al. (\cite{Juza94}) summarized the history of the investigation
of {\kdra} up to 1993.
They collected and homogenized all published records of spectral, light,
and polarimetric variations and combined them with their own
series of spectral and {\ubv} observations.
They found secular changes in $V$, $B$, {\bv}, and {\ub}, correlated
with changes of the strength of Balmer emission lines.
Notably enough, the $U$ magnitude only exhibited small, possibly cyclic
variations but no secular changes.

Juza et al. (\cite{Juza94}) analyzed all homogenized records of the
$E/C$ strength of the {\Hbeta} emission for periodicity and concluded
that it varies with a period of $8406\pm23$ days (23.01\,years).
From the records of the equivalent width {\WHalpha} they found the same
period of $8400\pm340$ days.

\begin{figure}
\resizebox{\hsize}{!}{\includegraphics{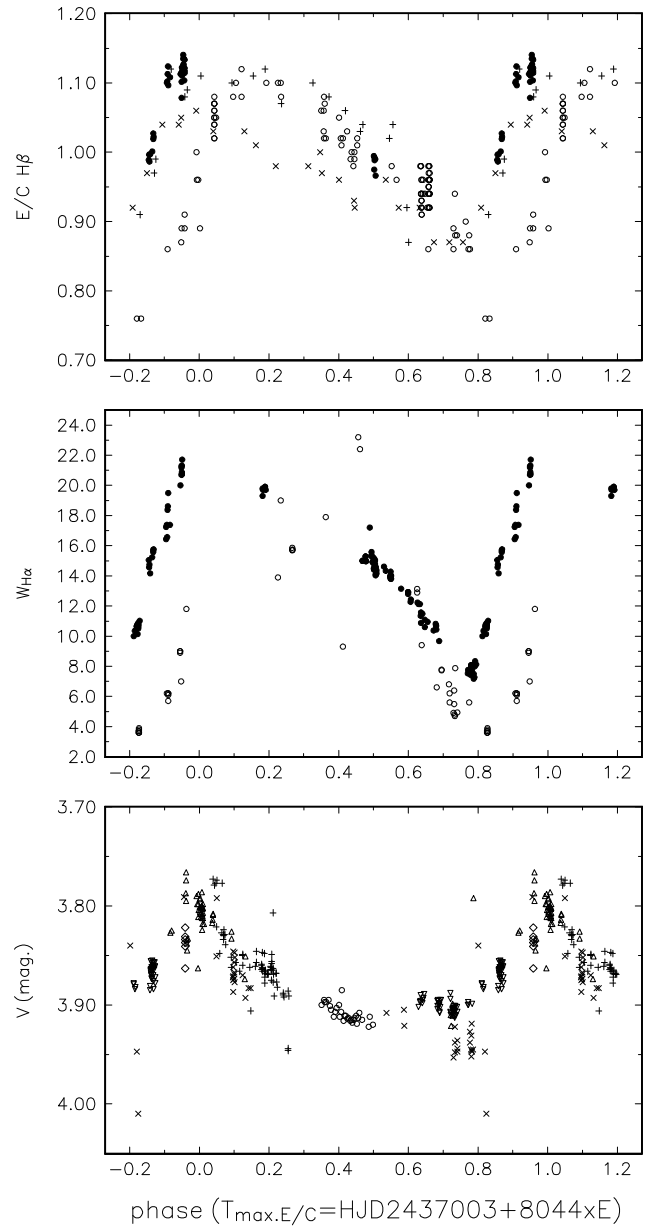}}
\caption{A phase plot of the {\Hbeta} $E/C$ emission strength,
{\WHalpha} and $V$ magnitude of {\kdra} vs. phase of the 8040-d period.
In the first panel, crosses denote the estimates compiled by
Jessup (JD~2412450-26680), pluses those by Curtiss (JD~2419546-25751),
open circles refer to measurements from photographic spectra published
by Juza et al. (JD~2441443-46587) and filled circles are
our measurements in electronic spectra (JD~49079-52754).
In the central panel, open and filled circles refer to equivalent width
measured in photographic and electronic spectra, respectively.
For $V$ photometry, the same notation is used as in Fig.~\ref{photo}.}
\label{8044}
\end{figure}

\subsection{Determination of a long term period}

Having new longer series of observations at our disposal, we repeated
the period analysis of Juza et al. (\cite{Juza94}) but with the
following modifications.

The principal data set, covering the longest time interval from
JD~2412450 to JD~2452754 is represented by various records of the
{\Hbeta} emission strength, $E/C$.
A proper period analysis of these observations is a bit tricky.
As we already mentioned, we prefer to analyze directly observable
quantities.
For $E/C$ we adopt the observed peak intensity of the emission,
measured in units of local continuum, i.e. $(I_V+I_R)/2$, similarly as
Juza et al. (\cite{Juza94}) did.
Older observations, based on visual estimates, were homogenized and
brought onto this scale already by Juza et al. but come from two
principal sources which we treat separately.
The third data set is represented by measurements from photographic
spectra and the last and the
most accurate one comes from electronic
spectra.
It is quite probable that the zero points of the $E/C$ scale for the
above mentioned four data subsets differ.
We allowed for it and the period analysis then showed that all data can
be reconciled quite
well with a period of ($8044\pm157$) days, the epoch
of the maximum strength being (HJD~$2437003\pm55$)\,d, i.e. 22.0\,years,
which is for about one year shorter than the previous estimate.

Then we kept this period fixed and analyzed {\WHalpha}, this time
allowing for a zero-point difference only between photographic and
electronic spectra. This data set spans a time interval
JD~2434437 -- 2452693.
Finally, we also folded all $V$ magnitude observations, spanning
JD~2435400 -- 2452827, with the same period.
The phase plot for all three observables is shown in Fig.~\ref{8044}.

The $E/C$ data now span full 5 cycles.
It is immediately seen that they can be folded with the 8044-d period to
the point that the maxima and minima are approximately presented but one
can also see the individuality of various cycles. For instance, the
latest cycle is distinctly narrower than the previous one.
Note that the period of 8400 days determined by Juza et al.
(\cite{Juza94}) predicts the the time of minimum {\WHalpha} for the year
2002 while the actual time of minimum came in 1999, i.e., three years
earlier.
Moreover, the time span between two latest events of a steep rise of the
{\Halpha} emission (cf. Fig.\,\ref{longvar}d) is $\sim6900$ days.
This puts serious doubts on a strict periodicity.

On the other hand, the mutual phase agreement of three {\sl independent}
data sets in Fig.~\ref{8044} looks quite impressive and there are
numerous examples among Be stars which display cyclic phenomena with
very variable individual cycle length and shape.
Therefore, the possibility of whether the whole phenomenon is not
controlled by some regular clock deserves further study.

\subsection{Discussion of long-term variations}

The question is how to interpret these spectacular long-term changes.
Already Juza et al. (\cite{Juza94}) pointed out that the character of
long-term variations of {\kdra} is typical for a positive correlation
between brightness and emission strength as defined by
Harmanec (\cite{hec83}).
They also noted that the rise of the emission strength is much steeper
than its decline.
From this fact, they concluded that there must be some, as yet unknown,
physical mechanism which supplies fresh material into the circumstellar
envelope every 23 years.

A phase lag between the maximum brightness and maximum emission strength
found by Juza et al. (\cite{Juza94}) has been confirmed.
This again agrees with the idea of two types of a correlation between
the brightness and emission strength (cf. Harmanec \cite{hec00}).
For an inverse correlation (opposite to that observed for {\kdra}, this
was discussed semi-quantitatively by Koubsk\'y et al. (\cite{4Her}):
When a new envelope is formed, it first adopts the form of a relatively
small region which is optically thick in the optical continuum (called
``pseudophotosphere'' by Harmanec \cite{hec83, hec00}).
If the star is observed more pole-on than equator-on (and this is the
case of {\kdra}), the inner parts of the disk effectively increase
the apparent radius of the star.
This leads to an increasing brightness of the object.
As the envelope grows in extent and rarefies, it becomes optically thin
in continuum and one observes increasingly strong Balmer emission and
the brightness begins to decrease again.

A very similar interpretation of the activity cycle was put forward
also by Hirata (\cite{talatabe}) who, however, modelled the optically
thick parts of the envelope explicitly as gravitationally darkened
stellar photospheres rotating close to the break-up speed.
We note that his approach is only legitimate if the envelope is formed
by a gradual outflow of the material from the stellar photosphere.

In principle, one can envision the following {\sl qualitative}
interpretations of the long-term changes observed:
\begin{itemize}
\item One possibility is that the long-term variations of {\kdra} are
manifestations of the still mysterious Be phenomenon.
Even so, the interesting aspect is that the re-appearance of the
emission is rather regular and that the envelope never disappears
completely as it is the case for many other Be stars.
If this explanation is accepted, {\kdra} is indeed a good example of
the object with a positive correlation between brightness and emission
strength.
\item Another possibility is to assume that the re-appearance of the
emission is a periodic phenomenon.
The fact that different cycles do not repeat perfectly need not be an
ultimate argument against such a possibility since there are many
examples of periodic phenomena with cycle-to-cycle variations, like
X-ray flares in Be+X binaries or even light curves of Be stars in
binary systems.
To the best of our knowledge, there are no speckle-interferometric
observations of {\kdra} and the idea that the formation of a new
envelope occurs during periastron passage of a putative companion to
{\kdra} moving in an eccentric orbit is a reasonable working hypothesis
to be tested by future observations.
\end{itemize}

\section{Summary}

This study is essentially devoted to the investigation of the long-term
variations and of the physical properties of the Be star {\kdra}.
We have presented and analysed new set of spectroscopic observations
obtained between 1992 -- 2003.
We also used the published records of the star.

Spectrograms obtained during last three years of observations cover a
wide range of the wavelengths (3450 -- 8860\,{\AA}).
These wide range spectra were used for a detailed line identification in
the visible and near infrared range and for a comparison with synthetic
spectra calculated from NLTE model atmospheres (Section \ref{modcomp}).
The best fit was found for a model with {\Teff}=14\,000K,
$\log g = 3.5$ and $v \sin i = 170\kms$.
However, it is only an approximate estimate, since we considered an
oversimplified picture of the atmosphere as a plane-parallel and static
one.
From the available photometric records, parallax, and from the model
atmosphere we obtained the stellar radius $R_\ast=6.4R_{\sun}$,
and mass $M_{\ast}=4.8M_{\sun}$.

An analysis of time variations of the line strength (equivalent width)
and intensity of {\Halpha} and {\Hbeta} over the last thirty years show
that {\kdra} undergoes cyclic long-term variations with an average cycle
of 22 years.
These cyclic changes are most probably caused by a steep rise and
a gradual decrease of the dimensions of a rapidly rotating circumstellar
disk.
This picture is supported by the fact that for stronger Balmer line
emission, formed in more distant (and not so fastly rotating) regions
of the disk, the peak separation gets also smaller.
This is observed both for individual Balmer emission lines at the same
spectrograms and for temporal evolution of each of the Balmer lines.

There remain unsolved problems, however.
First, the true physical origin of a 22 year variation cycle is unknown.
It is probably caused by disk oscillations, but that is all we are able
to say now.
Second, the fact that the helium lines are in absorption while hydrogen
and iron lines are in emission is also not explained.
More detailed modelling of a hydrodynamics and radiative transfer in a
disk will help to improve the basic physical parameters of this star.


\begin{acknowledgements}
Some of the {\Reticon} spectra from the archive of the Ond\v{r}ejov
2m-telescope (see \v{S}koda \cite{retarch}) have been secured by Drs.
J.~Horn, K.~Juza, V.~\v{S}imon, and S.~\v{S}tefl.
The authors would like to thank the technical staff of the 2m-telescope
(J.~Havelka, J.~Honsa, K.~Kala\v{s}, L.~\v{R}ezba, L.~\v{S}arounov\'a,
M.~Tlamicha, and F.~{\Zdarsky}) for their help with the observations.
Our thanks are also due to Dr. A.-M.~Hubert who kindly gave their
quantitative measurements of the {\Halpha} profiles at our disposal
and to Dr. Karen Bjorkman who kindly provided
us with a copy of an unpublished poster paper.
This research has made use of the NASA's Astrophysics Data System
Abstract Service (Kurtz et al. \cite{ADS1}, Eichhorn et al. \cite{ADS2},
Accomazzi et al. \cite{ADS3}, Grant et al. \cite{ADS4}) and the CDS
bibliopgraphical service.
Our work was supported by grants of Grant Agency of the Czech Republic
205/02/0445 and 205/03/0788.
Astronomical Institute Ond\v{r}ejov is supported by projects K2043105
and Z1003909.
Astronomical Institute of the Charles University is supported via
research plan J13/98: 113200004 of M\v{S}MT.
\end{acknowledgements}

\newcommand{\MSA}{in Modelling of Stellar Atmospheres, IAU Symp. 210,
 N. E. Piskunov, W. W. Weiss, \& D. F. Gray eds., ASP Conf. Ser.,
 in press}

\newcommand{\Alicante}[1]{in The Be Phenomenon in Early Type Stars, IAU
 Coll. 175, M. A. Smith, H. F. Henrichs, \& J. Fabregat eds., ASP
 Conf. Ser. Vol. 214, p. #1}


\Online

In all tables, $W$ is the equivalent width, $I_c$ is the central line
intensity, $I_V$ and $I_R$ are the violet and red emission peak
intensities, respectively, {\Delnup} is the velocity difference between
red and violet peaks, and $V/R$ is the violet-to-red peak ratio.
The sign --- means that the corresponding quantity was not
measureable.

\begin{table*}
\caption{Measured parameters of the {\Halpha} emission line.
}\label{Hatable}
\begin{tabular}{|l|l|cccccc|} \hline
\hline
\textbf{File}     &
\textbf{HJD}      &
\textbf{\WHalpha} &
\textbf{$I_c$}    &
\textbf{$I_V$}    &
\textbf{$I_R$}    &
\textbf{\Delnup}  &
\textbf{$V/R$}    \\
& $-$2400000 & [\AA] & & & & [\kms] & \\ \hline

 R00311 & 48813.4315& $-14.99$ & 2.68& 2.95& 2.77& 124.5& 1.06\\
 R00685 & 48883.5010& $-15.27$ & 2.80& 2.96& 2.86& 107.6& 1.03\\
 R00743 & 48893.2294& $-14.96$ & 2.73& 2.86& 2.88& 93.05& 0.99\\ 
 R00939 & 49018.5005& $-15.59$ & 2.75& 2.94& 3.05& 108.3& 0.96\\
 R00940 & 49018.5747& $-15.38$ & 2.72& 2.87& 2.97& 106.0& 0.96\\
 R01270 & 49045.4875& $-14.92$ & 2.70& 2.86& 2.84& 97.70& 1.00\\
 R01303 & 49066.4515& $-15.15$ & 2.72& 2.96& 2.81& 114.1& 1.05\\
 R01381 & 49079.5350& $-14.46$ & 2.68& 2.77& 2.83& 115.9& 0.97\\
 R01448 & 49081.4524& $-14.41$ & 2.68& 2.80& 2.85& 113.0& 0.98\\
 R01465 & 49088.3750& $-14.96$ & 2.75& 2.85& 2.87& 109.1& 0.99\\
 R01515 & 49092.4614& $-14.71$ & 2.72& 2.84& 2.84& 125.3& 0.99\\
 R01566 & 49102.3636& $-14.78$ & 2.72& 2.80& 2.85& 110.1& 0.98\\
 R01568 & 49102.3818& $-15.07$ & 2.75& 2.83& 2.87& 108.3& 0.98\\
 R01697 & 49116.4731& $-14.03$ & 2.58& 2.90& 2.72& 129.9& 1.06\\
 R01740 & 49119.5657& $-14.57$ & 2.63& 3.05& 2.78& 124.1& 1.09\\
 R01839 & 49133.3452& $-14.12$ & 2.68& 2.78& 2.74& 120.7& 1.01\\
 R01849 & 49133.4032& $-14.30$ & 2.71& 2.82& 2.79& 117.7& 1.00\\
 R03467 & 49310.6279& $-14.61$ & 2.67& --- & 2.75& 111.3& --- \\
 R03601 & 49350.6331& $-14.34$ & 2.68& 2.79& 2.81& 103.6& 0.99\\
 R04243 & 49463.3294& $-14.28$ & 2.66& 2.73& 2.81& 125.9& 0.97\\
 R04249 & 49463.3564& $-13.99$ & 2.64& 2.71& 2.78& 127.1& 0.97\\
 R04261 & 49463.4929& $-14.04$ & 2.65& 2.72& 2.79& 125.9& 0.97\\
 R04274 & 49463.5259& $-13.90$ & 2.64& 2.72& 2.78& 125.9& 0.97\\
 R04406 & 49466.3311& $-13.98$ & 2.62& 2.74& 2.79& 126.5& 0.98\\
 R04407 & 49466.3459& $-13.86$ & 2.62& 2.74& 2.77& 123.6& 0.98\\
 R04427 & 49466.3994& $-13.88$ & 2.62& 2.74& 2.77& 126.5& 0.98\\
 R04433 & 49466.4142& $-13.83$ & 2.62& 2.74& 2.77& 126.5& 0.99\\
 R04445 & 49466.5338& $-13.95$ & 2.62& 2.75& 2.79& 127.1& 0.98\\
 R04446 & 49466.5492& $-13.97$ & 2.63& 2.74& 2.79& 126.6& 0.98\\
 R04462 & 49466.6149& $-13.96$ & 2.62& 2.75& 2.79& 126.5& 0.98\\
 R04469 & 49467.3034& $-14.03$ & 2.64& 2.74& 2.77& 124.2& 0.99\\
 R04484 & 49467.3497& $-13.98$ & 2.65& 2.74& 2.77& 125.3& 0.99\\
 R04496 & 49467.4651& $-13.79$ & 2.63& 2.72& 2.74& 124.8& 0.99\\
 R04515 & 49467.5566& $-13.82$ & 2.63& 2.73& 2.75& 125.4& 0.99\\
 R07348 & 49702.6550& $-13.14$ & 2.55& 2.64& 2.73& 121.8& 0.96\\
 R08190 & 49862.3759& $-12.80$ & 2.43& 2.74& 2.60& 128.3& 1.05\\
 R08191 & 49862.3842& $-12.92$ & 2.44& 2.76& 2.61& 130.1& 1.05\\
 R08551 & 49918.3372& $-12.28$ & 2.37& 2.71& 2.49& 140.6& 1.09\\
 R08739 & 49930.3387& $-12.42$ & 2.45& 2.66& 2.52& 130.1& 1.06\\
 R09690 & 50080.6261& $-12.23$ & 2.42& 2.49& 2.63& 123.0& 0.95\\
 R09862 & 50104.7034& $-12.13$ & 2.37& 2.67& 2.49& 135.9& 1.07\\
 R10089 & 50140.5900& $-12.12$ & 2.41& 2.49& 2.58& 119.5& 0.96\\
 R10245 & 50158.5779& $-11.35$ & 2.35& 2.55& 2.38& 126.9& 1.07\\
 R10268 & 50159.4882& $-10.87$ & 2.27& 2.51& 2.33& 133.4& 1.07\\
 R10285 & 50160.4201& $-11.57$ & 2.35& 2.59& 2.43& 133.9& 1.06\\
 R10562 & 50193.4706& $-11.49$ & 2.34& 2.46& 2.55& 132.2& 0.96\\
 R10972 & 50249.4667& $-10.60$ & 2.22& 2.38& 2.39& 124.6& 0.99\\
 R10995 & 50251.3830& $-11.08$ & 2.29& 2.46& 2.46& 135.8& 1.00\\
 R11509 & 50316.3418& $-10.95$ & 2.29& 2.38& 2.36& 109.4& 1.00\\
 R12257 & 50448.6371& $-10.38$ & 2.20& 2.28& 2.39& 126.4& 0.95\\
 R12528 & 50497.4534& $-10.82$ & 2.24& 2.44& 2.48& 122.8& 0.98\\
 R12576 & 50506.5231& $-10.71$ & 2.27& 2.35& 2.44& 128.5& 0.96\\
 R12619 & 50509.529 & $-10.66$ & 2.23& 2.33& 2.44& 125.7& 0.95\\
 R12743 & 50518.464 & $-10.44$ & 2.30& 2.39& 2.36& 135.2& 1.01\\
\hline
\end{tabular}
\end{table*}

\addtocounter{table}{-1}
\begin{table*}
\caption{ -- continued}
\begin{tabular}{|l|l|cccccc|} \hline
\hline
\textbf{File}     &
\textbf{HJD}      &
\textbf{\WHalpha} &
\textbf{$I_c$}    &
\textbf{$I_V$}    &
\textbf{$I_R$}    &
\textbf{\Delnup}  &
\textbf{$V/R$}    \\
& $-$2400000 & [\AA] & & & & [\kms] & \\ \hline

 R13044 & 50583.4009& $ -9.676$& 2.14& 2.33& 2.28& 136.8& 1.02\\
 R13544 & 51238.5568& $ -7.531$& 1.87& 1.94& 1.99& 131.7& 0.97\\
 R13622 & 51250.4931& $ -7.761$& 1.87& 1.94& 2.04& 122.3& 0.95\\
 R13741 & 51304.5682& $ -7.592$& 1.88& 1.93& 1.99& 131.1& 0.96\\
 R13795 & 51316.3680& $ -7.629$& 1.88& 1.95& 2.00& 126.4& 0.97\\
 R13799 & 51316.4734& $ -7.415$& 1.83& 1.90& 1.94& 126.4& 0.98\\
 R13849 & 51323.3312& $ -8.020$& 1.89& 2.04& 2.00& 128.7& 1.02\\
 R13851 & 51323.3469& $ -8.040$& 1.89& 2.04& 2.00& 132.8& 1.02\\
 R13853 & 51323.3652& $ -8.020$& 1.89& 2.04& 2.00& 129.9& 1.02\\
 R13855 & 51323.3851& $ -8.095$& 1.89& 2.04& 2.00& 128.7& 1.02\\
 R13857 & 51323.4095& $ -8.045$& 1.89& 2.04& 2.00& 129.3& 1.02\\
 R13859 & 51323.4402& $ -8.078$& 1.89& 2.04& 2.00& 128.7& 1.02\\
 R13868 & 51325.3803& $ -7.913$& 1.90& 2.05& 1.95& 127.5& 1.05\\
 R13920 & 51328.4117& $ -7.700$& 1.79& 2.05& 1.93& 136.9& 1.06\\
 R13922 & 51328.4401& $ -7.790$& 1.81& 2.05& 1.94& 139.2& 1.05\\
 R14118 & 51378.5441& $ -7.170$& 1.83& 1.91& 1.90& 142.7& 1.01\\
 R14137 & 51379.4291& $ -7.501$& 1.86& 1.94& 1.92& 119.3& 1.01\\
 R14276 & 51391.4448& $ -7.260$& 1.78& 1.98& 1.87& 110.0& 1.06\\
 R14438 & 51401.3442& $ -7.992$& 1.85& 2.01& 1.88& 138.2& 1.07\\
 R14477 & 51410.3629& $ -8.344$& 1.96& 2.06& 1.96& 119.9& 1.05\\
 R14787 & 51433.3458& $ -8.124$& 1.95& 2.00& 2.00& 104.1& 1.00\\
 R15597 & 51580.6094& $-10.00$ & 2.11& 2.31& 2.13& 131.0& 1.08\\
 R15740 & 51602.4666& $-10.35$ & 2.19& 2.24& 2.30& 107.5& 0.97\\
 R15898 & 51643.4587& $-10.65$ & 2.19& 2.38& 2.25& 126.4& 1.05\\
 R15973 & 51656.3361& $-10.60$ & 2.24& 2.32& 2.31& 117.0& 1.00\\
 R16003 & 51661.4248& $-10.77$ & 2.25& 2.32& 2.33& 108.2& 0.99\\
 R16079 & 51669.3846& $-10.13$ & 2.19& 2.28& 2.26& 114.1& 1.01\\
 R16137 & 51678.3676& $-10.76$ & 2.25& 2.32& 2.35& 113.5& 0.98\\
 R16148 & 51679.3875& $-10.72$ & 2.25& 2.32& 2.33& 117.6& 0.99\\
 R16150 & 51679.4007& $-10.73$ & 2.25& 2.32& 2.34& 117.6& 0.99\\
 R16153 & 51679.4206& $-10.79$ & 2.25& 2.33& 2.34& 118.2& 0.99\\
 R16154 & 51679.4386& $-10.88$ & 2.26& 2.34& 2.35& 119.3& 0.99\\
 R16167 & 51680.3668& $-10.61$ & 2.25& 2.31& 2.31& 118.8& 0.99\\
 R16195 & 51681.4057& $-10.51$ & 2.24& 2.29& 2.29& 118.8& 1.00\\
 R16505 & 51714.3660& $-11.01$ & 2.26& 2.40& 2.34& 117.6& 1.02\\

 HR1349 & 51924.5409& $-15.04$ & 2.78& 2.93& 2.90& 108.0& 1.01\\
 HR1396 & 51936.5148& $-14.57$ & 2.75& 2.90& 2.89& 97.28& 1.00\\
 HR1417 & 51938.5012& $-14.74$ & 2.72& 2.92& 2.87& 84.73& 1.02\\
 HR1613 & 51959.5535& $-14.16$ & 2.65& 2.89& 2.74& 81.30& 1.06\\
 HR1852 & 52005.3418& $-15.23$ & 2.83& 3.01& 2.93& 68.74& 1.03\\
 HR1951 & 52027.4415& $-15.57$ & 2.86& 2.99& --- & 97.30& ---\\
 HR1966 & 52029.3652& $-15.70$ & 2.87& 3.01& --- & 82.44& ---\\
 HR2058 & 52038.5645& $-15.76$ & 2.85& --- & --- & 109.8& ---\\
 CCD1639& 52322.6262& $-16.43$ & 3.01& 3.20& 2.95& 91.44& 1.08\\
 CCD1640& 52322.6737& $-17.24$ & 3.03& 3.30& 2.92& 95.55& 1.13\\
 HR3339 & 52343.4912& $-17.38$ & 3.04& 3.15& 3.14& 80.53& 1.00\\
 HR3465 & 52352.5557& $-16.57$ & 3.08& 3.24& --- & 85.42& ---\\
 HR3492 & 52362.3688& $-18.37$ & --- & --- & --- & 72.17& ---\\
 HR3567 & 52366.3847& $-18.62$ & 3.24& --- & --- & 64.43& ---\\
 HR3658 & 52373.4362& $-19.50$ & 3.31& 3.48& 3.44& 75.60& 1.01\\
 HR3977 & 52417.4422& $-17.38$ & 2.95& 3.35& --- & 75.60& ---\\
 HR5335 & 52657.5959& $-20.00$ & 3.46& 3.46& 3.54& 79.48& 0.97\\
 HR5411 & 52683.5671& $-21.27$ & 3.56& 3.63& 3.60& 56.41& 1.01\\
 HR5431 & 52684.4712& $-21.17$ & 3.62& 3.67& 3.64& 56.19& 1.00\\
 HR5434 & 52684.5418& $-20.70$ & 3.57& 3.60& 3.59& 53.67& 1.00\\
 HR5481 & 52687.4981& $-20.72$ & 3.52& --- & --- & ---  & --- \\
 HR5484 & 52687.5881& $-20.76$ & --- & --- & --- & ---  & --- \\
 HR5495 & 52688.4042& $-21.31$ & --- & 3.71& --- & ---  & --- \\
 HR5524 & 52692.4424& $-20.87$ & 3.41& 3.62& 3.38& 94.10& 1.07\\
 HR5537 & 52693.4215& $-21.70$ & 3.49& 3.74& 3.48& 94.33& 1.07\\
\hline
\end{tabular}
\end{table*}

\addtocounter{table}{-1}
\begin{table*}
\caption{ -- continued}
\begin{tabular}{|l|l|cccccc|} \hline
\hline
\textbf{File}     &
\textbf{HJD}      &
\textbf{\WHalpha} &
\textbf{$I_c$}    &
\textbf{$I_V$}    &
\textbf{$I_R$}    &
\textbf{\Delnup}  &
\textbf{$V/R$}    \\
& $-$2400000 & [\AA] & & & & [\kms] & \\ \hline

 HR5757 & 52720.3170& $-20.75$ & 3.41& 3.54& 3.52& 91.13& 1.01\\
 HR5793 & 52721.4772& $-20.78$ & 3.50& --- & --- & 79.56& ---\\
 HR5816 & 52722.5142& $-21.69$ & 3.58& 3.75& 3.62& 86.56& 1.04\\
 HR5836 & 52723.4036& $-20.56$ & 3.27& 3.33& --- & 76.51& --- \\
 HR5837 & 52723.4325& $-21.07$ & 3.49& 3.62& 3.54& 88.62& 1.02\\
 HR5864 & 52724.4549& $-20.74$ & 3.49& 3.66& 3.52& 96.38& 1.04\\
 HR5926 & 52727.4657& $-21.50$ & 3.49& 3.63& 3.53& 84.51& 1.03\\
 md0412 & 52734.7316& $-20.51$ & 3.39& 3.37& 3.43& 64.20& 0.984\\
 md1118 & 52741.9524& $-19.92$ & 3.33& 3.47& 3.36& 63.89& 1.032\\
 md1221 & 52742.6241& $-20.37$ & 3.38& 3.53& 3.42& 60.96& 1.032\\
 md1222 & 52742.6307& $-20.47$ & 3.37& 3.53& 3.45& 60.37& 1.024\\
 md1421 & 52744.8238& $-20.44$ & 3.42& 3.56& 3.45& 52.75& 1.032\\
 md1422 & 52744.3828& $-19.89$ & 3.39& 3.51& 3.41& 53.34& 1.030\\
 md1426 & 52744.8590& $-20.06$ & 3.40& 3.52& 3.43& 53.34& 1.026\\
 md2111 & 52751.7782& $-19.71$ & 3.21& 3.54& 3.25& 114.8& 1.087\\
 md2112 & 52751.7863& $-19.81$ & 3.22& 3.55& 3.26& 115.5& 1.091\\
 md2114 & 52751.7979& $-19.83$ & 3.23& 3.55& 3.25& 115.5& 1.09\\
 md2115 & 52751.8014& $-19.78$ & 3.21& 3.54& 3.26& 114.9& 1.088\\
 md2229 & 52752.5113& $-19.79$ & 3.23& 3.53& 3.27& 113.1& 1.078\\
 md2429 & 52754.8454& $-19.97$ & 3.20& 3.53& 3.30& 113.1& 1.07\\
 md2430 & 52754.8493& $-20.09$ & 3.19& 3.53& 3.30& 111.3& 1.07\\
 md2431 & 52754.8534& $-20.21$ & 3.21& 3.54& 3.32& 113.7& 1.07\\
 md2432 & 52754.8576& $-19.96$ & 3.19& 3.52& 3.32& 113.1& 1.062\\
 md2433 & 52754.8600& $-20.08$ & 3.22& 3.56& 3.29& 111.9& 1.081\\
 md2434 & 52754.8624& $-19.93$ & 3.20& 3.54& 3.29& 111.4& 1.077\\
 md2435 & 52754.8644& $-20.12$ & 3.21& 3.54& 3.30& 111.9& 1.07\\
 md2436 & 52754.8680& $-19.98$ & 3.20& 3.52& 3.31& 111.4& 1.063\\
 md2437 & 52754.8699& $-19.97$ & 3.20& 3.54& 3.31& 110.8& 1.068\\
 md2438 & 52754.8733& $-19.89$ & 3.22& 3.53& 3.30& 112.5& 1.07\\
\hline
\end{tabular}
\end{table*}

\begin{table*}
\caption{Measured parameters of {\Hbeta}, {\Hgamma}, and {\Hdelta}
lines.}\label{Balpar}
\scalebox{.90}{\includegraphics{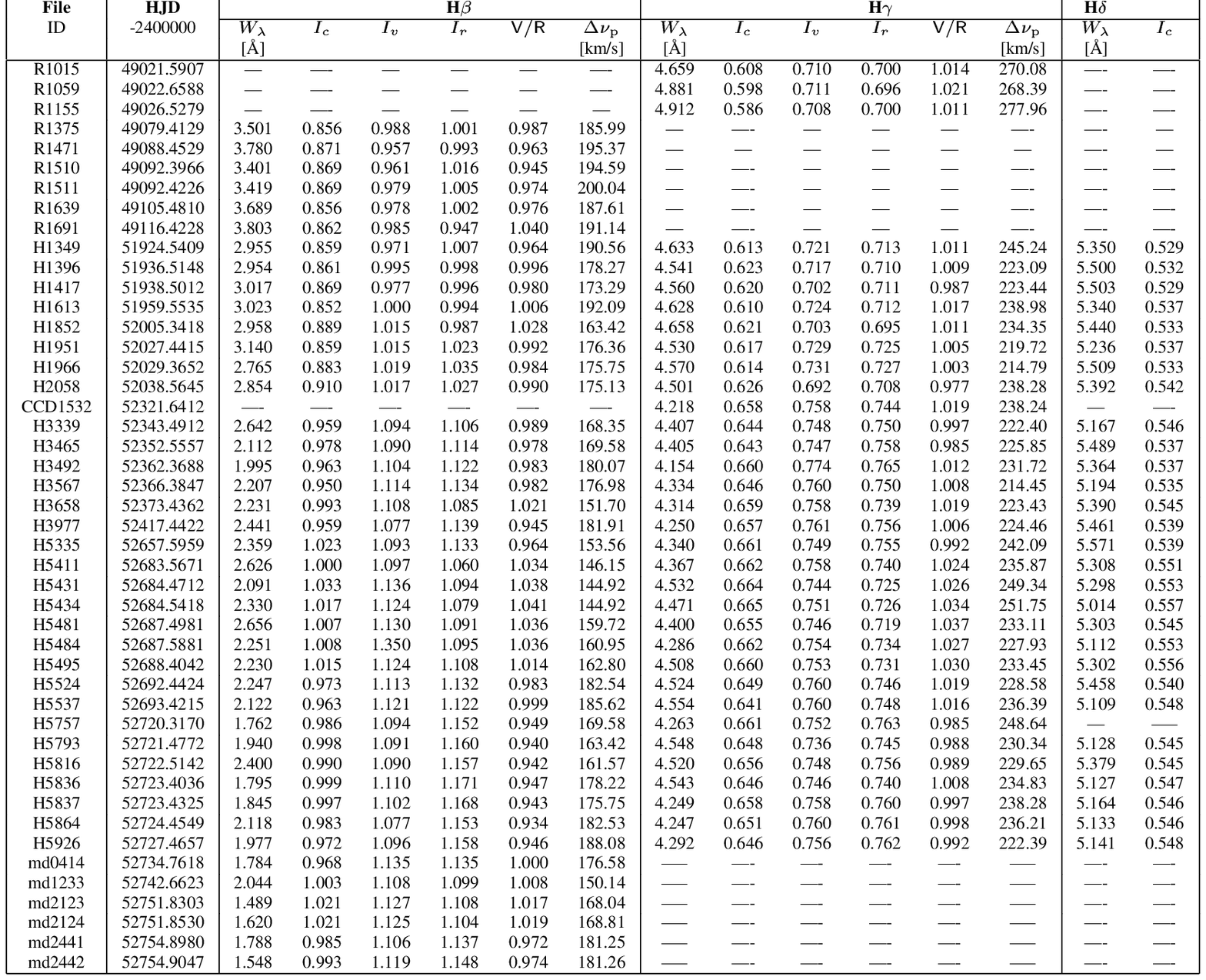}}
\end{table*}

\begin{table*}
\caption{Measured parameters of {\ion{Si}{ii} 6347}, {\ion{Si}{ii}
6371}, {\ion{He}{i} 6678}, and \ion{Fe}{ii} 6456
lines.}\label{hefesi}
\begin{tabular}{|l|l|cc|cc|cc|cc|} \hline \hline
\textsf{File} & \textsf{HJD}&
\multicolumn{2}{|c|}{\textsf{\ion{Si}{ii} 6347\,{\AA}}} &
\multicolumn{2}{|c|}{\textsf{\ion{Si}{ii} 6371\,{\AA}}} &
\multicolumn{2}{|c|}{\textsf{\ion{He}{i}  6678\,{\AA}}} &
\multicolumn{2}{|c|}{\textsf{\ion{Fe}{ii} 6456\,{\AA}}} \\
\cline {3-10}
& $-2400000$ & $W[\mbox{\AA}]$ & $I_c$
             & $W[\mbox{\AA}]$ & $I_c$
             & $W[\mbox{\AA}]$ & $I_c$
             & $W[\mbox{\AA}]$ & $I_c$ \\
\hline
R00311 &
48813.4315 & 0.04& 0.970& 0.001& 0.981& 0.172& 0.971& $-0.166$& 1.019\\
R00685 &
48883.50105& 0.11& 0.957& 0.007& 0.983& 0.150& 0.978& $-0.089$& 1.010\\
R00743 &
48893.2294 & 0.13& 0.953& 0.077& 0.974& 0.206& 0.970& $-0.052$& 1.007\\
R00940 &
49018.5747 &  ---&  --- &  --- &  --- & 0.166& 0.977& $-0.165$& 1.015\\
R01270 &
49045.4875 & 0.14& 0.969& 0.130& 0.976& 0.201& 0.967& $-0.102$& 1.006\\
R01303 &
49066.4515 & 0.14& 0.964& 0.131& 0.972& 0.197& 0.964& $-0.044$& 1.001\\
R01381 &
49079.5350 & 0.13& 0.964& 0.112& 0.972& 0.213& 0.965& $-0.074$& 1.005\\
R01442 &
49081.4162 & 0.12& 0.970& 0.034& 0.983& 0.204& 0.963& $-0.162$& 1.012\\
R01448 &
49081.4524 & 0.14& 0.961& 0.083& 0.981& 0.205& 0.953& $-0.048$& 0.999\\
R01465 &
49088.3750 & 0.08& 0.976&  --- &  --- & 0.203& 0.957& $-0.098$& 1.008\\
R01515 &
49092.4614 & 0.07& 0.972&  --- &  --- & 0.202& 0.960& $-0.074$& 1.007\\
R01565 &
49102.3385 & 0.09& 0.973&  --- &  --- & 0.170& 0.968& $-0.068$& 1.003\\
R01566 &
49102.3636 & 0.07& 0.977&  --- &  --- & 0.161& 0.969& $-0.096$& 1.004\\
R01568 &
49102.3818 & 0.07& 0.977&  --- &  --- & 0.191& 0.966& $-0.103$& 1.009\\
R01697 &
49116.4731 & 0.12& 0.957& 0.102& 0.979& 0.209& 0.955&   ---   &  --- \\
R01740 &
49119.5657 & 0.08& 0.971& 0.061& 0.978& 0.207& 0.964&   ---   &  --- \\
R01839 &
49133.3452 & 0.13& 0.966& 0.137& 0.964& 0.162& 0.975&   ---   &  --- \\
R01849 &
49133.4032 & 0.13& 0.957& 0.054& 0.978& 0.157& 0.967&   ---   &  --- \\
R03466 &
49310.5844 & 0.08&  --- & 0.026&  --- & 0.182&  --- & $-0.155$&  --- \\
R03467 &
49310.6279 & 0.13&  --- & 0.038&  --- & 0.164&  --- & $-0.112$&  --- \\
R03601 &
49350.6331 & 0.08& 0.974& 0.095& 0.978& 0.164& 0.969& $-0.171$& 1.015\\
R03848 &
49403.6115 & 0.08& 0.975& 0.078& 0.982& 0.136& 0.973& $-0.116$& 1.009\\
R04243 &
49463.3294 & 0.08& 0.966& 0.074& 0.974& 0.168& 0.967& $-0.143$& 1.011\\
R04249 &
49463.3564 & 0.09& 0.963& 0.074& 0.975& 0.175& 0.965& $-0.098$& 1.008\\
R04261 &
49463.4929 & 0.09& 0.966& 0.089& 0.978& 0.160& 0.968& $-0.128$& 1.014\\
R04274 &
49463.5259 & 0.11& 0.966& 0.093& 0.975& 0.175& 0.966& $-0.086$& 1.009\\
R04406 &
49466.3311 & 0.09& 0.970& 0.081& 0.979& 0.171& 0.972& $-0.119$& 1.011\\
R04407 &
49466.3459 & 0.10& 0.965& 0.077& 0.976& 0.172& 0.969& $-0.074$& 1.010\\
R04427 &
49466.3994 & 0.10& 0.970& 0.078& 0.978& 0.127& 0.971& $-0.138$& 1.017\\
R04433 &
49466.4142 & 0.10& 0.966& 0.072& 0.981& 0.167& 0.971& $-0.079$& 1.011\\
R04445 &
49466.5338 & 0.09& 0.963& 0.074& 0.980& 0.162& 0.969& $-0.133$& 1.011\\
R04446 &
49466.5492 & 0.09& 0.964& 0.079& 0.976& 0.162& 0.967& $-0.076$& 1.010\\
R04462 &
49466.6149 & 0.09& 0.962& 0.080& 0.973& 0.167& 0.965& $-0.131$& 1.011\\
R04469 &
49467.3034 & 0.10& 0.964& 0.077& 0.976& 0.185& 0.964& $-0.076$& 1.007\\
R04484 &
49467.3497 & 0.10& 0.963& 0.082& 0.977& 0.171& 0.969& $-0.130$& 1.013\\
R04496 &
49467.4651 & 0.11& 0.967& 0.079& 0.979& 0.174& 0.968& $-0.083$& 1.010\\
R04515 &
49467.5566 & 0.11& 0.963& 0.082& 0.976& 0.182& 0.969& $-0.138$& 1.011\\
R07348 &
49702.6550 & 0.09& 0.982& 0.069& 0.984& 0.215& 0.969& $-0.145$& 1.011\\
R08124 &
49853.4086 & 0.09& 0.971& 0.074& 0.981& 0.188& 0.964& $-0.104$& 1.007\\
R08126 &
49853.4793 & 0.09& 0.972& 0.070& 0.979& 0.169& 0.968& $-0.031$& 1.001\\
R08190 &
49862.3759 & 0.09& 0.968& 0.085& 0.976& 0.191& 0.967& $-0.070$& 1.010\\
R08191 &
49862.3842 & 0.10& 0.975& 0.092& 0.974& 0.194& 0.967& $-0.033$& 1.003\\
R08551 &
49918.3372 & 0.09& 0.957& 0.084& 0.970& 0.205& 0.964& $-0.024$& 1.008\\
R08739 &
49930.3387 & 0.09& 0.966& 0.098& 0.974& 0.174& 0.971&   ---   &  --- \\
R09690 &
50080.6261 & 0.12& 0.970& 0.082& 0.980& 0.172& 0.966& $-0.071$& 1.006\\
R09862 &
50104.7034 & 0.11& 0.972& 0.088& 0.979& 0.172& 0.970&   ---   &  --- \\
R10089 &
50140.5900 & 0.10& 0.971& 0.091& 0.979& 0.173& 0.969&   ---   &  --- \\
R10245 &
50158.5779 & 0.10& 0.976& 0.080& 0.982& 0.183& 0.966&   ---   &  --- \\
R10268 &
50159.4882 & 0.08& 0.963& 0.122& 0.976& 0.187& 0.966&   ---   &  --- \\
R10285 &
50160.4201 & 0.09& 0.967& 0.097& 0.978& 0.187& 0.972&   ---   &  --- \\
R10562 &
50193.4706 & 0.10& 0.975& 0.125& 0.972& 0.182& 0.966& $-0.056$& 1.004\\
R10972 &
50249.4667 & 0.09& 0.969& 0.079& 0.975& 0.161& 0.968&   ---   &  --- \\
R10995 &
50251.3830 & 0.11& 0.958& 0.146& 0.963& 0.186& 0.974&   ---   &  --- \\
R11509 &
50316.3418 & 0.11& 0.960& 0.086& 0.982& 0.168& 0.963&   ---   &  --- \\
R12257 &
50448.6371 & 0.09& 0.973& 0.081& 0.981& 0.163& 0.970&   ---   &  --- \\
R12528 &
50506.5231 & 0.09& 0.970& 0.099& 0.976& 0.178& 0.967&   ---   &  --- \\
R12619 &
50509.5290 & 0.10& 0.970& 0.071& 0.981& 0.184& 0.967&   ---   &  --- \\
R12743 &
50518.4640 & 0.10& 0.967& 0.098& 0.975& 0.179& 0.966&   ---   &  --- \\
\hline
\end{tabular}
\end{table*}

\addtocounter{table}{-1}
\begin{table*}
\caption{ -- continued}
\begin{tabular}{|l|l|cc|cc|cc|cc|} \hline \hline
\textsf{File} & \textsf{HJD}&
\multicolumn{2}{|c|}{\textsf{\ion{Si}{ii} 6347\,{\AA}}} &
\multicolumn{2}{|c|}{\textsf{\ion{Si}{ii} 6371\,{\AA}}} &
\multicolumn{2}{|c|}{\textsf{\ion{He}{i}  6678\,{\AA}}} &
\multicolumn{2}{|c|}{\textsf{\ion{Fe}{ii} 6456\,{\AA}}} \\
\cline {3-10}
& $-2400000$ & $W[\mbox{\AA}]$ & $I_c$
           & $W[\mbox{\AA}]$ & $I_c$
           & $W[\mbox{\AA}]$ & $I_c$
           & $W[\mbox{\AA}]$ & $I_c$ \\
\hline
R13044 &
50583.4009 & 0.09& 0.969& 0.085& 0.974& 0.181& 0.965&   ---   &  --- \\
R13544 &
51238.5568 & 0.13& 0.978& 0.043& 0.988& 0.120& 0.978& $-0.105$& 0.978\\
R13622 &
51250.4931 & 0.03& 0.986& 0.106& 0.984& 0.122& 0.976& $-0.118$& 0.976\\
R13795 &
51316.3680 & 0.15& 0.975& 0.076& 0.977& 0.164& 0.977&    00000& 1.021\\
R13799 &
51316.4734 & 0.09& 0.954&  --- &  --- & 0.208& 0.960& $-0.079$& 1.021\\
R13842 &
51322.4657 & 0.09& 0.984& 0.050& 0.983& 0.152& 0.971&   ---   &  --- \\
R13849 &
51323.3312 & 0.10& 0.989& 0.043& 0.988& 0.144& 0.975& $-0.037$& 0.998\\
R13851 &
51323.3469 & 0.10& 0.981& 0.033& 0.995& 0.131& 0.974& $-0.133$& 1.010\\
R13853 &
51323.3652 & 0.07& 0.969& 0.039& 0.988& 0.151& 0.974& $-0.135$& 1.008\\
R13855 &
51323.3851 & 0.07& 0.973& 0.063& 0.981& 0.177& 0.966& $-0.175$& 1.014\\
R13857 &
51323.4095 & 0.07& 0.972& 0.042& 0.991& 0.165& 0.968& $-0.182$& 1.013\\
R13859 &
51323.4402 & 0.07& 0.974& 0.052& 0.984& 0.153& 0.969& $-0.147$& 1.009\\
R13868 &
51325.3803 & 0.08& 0.969& 0.066& 0.979& 0.144& 0.973& $-0.133$& 1.002\\
R13920 &
51328.4117 & 0.08& 0.963& 0.080& 0.968& 0.147& 0.972&   ---   &  --- \\
R13922 &
51328.4401 & 0.08& 0.962& 0.093& 0.972& 0.148& 0.973&   ---   &  --- \\
R14118 &
51378.5441 & 0.05& 0.966& 0.062& 0.979& 0.165& 0.974&   ---   &  --- \\
R14137 &
51379.4291 & 0.06& 0.969& 0.052& 0.983& 0.187& 0.974&   ---   &  --- \\
R14276 &
51391.4448 & 0.06& 0.970& 0.058& 0.978& 0.182& 0.972&   ---   &  --- \\
R14438 &
51401.3442 & --- &  --- &  --- &  --- & 0.114& 0.980&   ---   &  --- \\
R14477 &
51410.3629 & 0.05& 0.974& 0.049& 0.986& 0.148& 0.975&   ---   &  --- \\
R14787 &
51433.3458 & 0.07& 0.965& 0.083& 0.977& 0.158& 0.972&   ---   &  --- \\
R15597 &
51580.6094 & 0.06& 0.975&  --- &  --- & 0.133& 0.973& $-0.071$& 1.001\\
R15740 &
51602.4666 & 0.06& 0.983& 0.030& 0.990& 0.138& 0.975& $-0.181$& 1.011\\
R15898 &
51643.4587 & 0.04& 0.983& ---  &  --- & 0.112& 0.977& $-0.102$& 1.001\\
R15973 &
51656.3361 & 0.06& 0.974& 0.088& 0.974& 0.134& 0.975& $-0.100$& 0.996\\
R16003 &
51661.4248 & 0.05& 0.978& 0.039& 0.986& 0.130& 0.974& $-0.068$& 0.997\\
R16079 &
51669.3846 & 0.06& 0.969& 0.036& 0.983& 0.132& 0.976& $-0.127$& 1.011\\
R16137 &
51678.3676 & 0.05& 0.983& 0.026& 0.990& 0.139& 0.976& $-0.080$& 1.004\\
R16148 &
51679.3875 & 0.04& 0.976& 0.039& 0.988& 0.152& 0.975& $-0.213$& 1.010\\
R16150 &
51679.4007 & 0.05& 0.978& 0.035& 0.989& 0.127& 0.979& $-0.159$& 1.009\\
R16153 &
51679.4206 & 0.05& 0.979& 0.047& 0.984& 0.140& 0.974& $-0.223$& 1.017\\
R16154 &
51679.4386 & 0.05& 0.975& 0.038& 0.944& 0.142& 0.976& $-0.228$& 1.019\\
R16167 &
51680.3668 & 0.04& 0.975& 0.050& 0.984& 0.133& 0.972& $-0.211$& 1.015\\
R16195 &
51681.4057 & 0.05& 0.974& 0.042& 0.978& 0.140& 0.978& $-0.207$& 1.015\\
R16505 &
51714.3660 & 0.07& 0.971& 0.036& 0.985& 0.140& 0.975& $-0.085$& 1.002\\
HR1349&
51924.5409 & 0.10& 0.974& 0.052& 0.983& 0.170& 0.967& $-0.110$& 0.992\\
HR1396&
51936.5148 & 0.14& 0.959&  --- &  --- & 0.114& 0.977& $-0.253$& 1.009\\
HR1417&
51938.5012 & 0.12& 0.966&  --- &  --- & 0.194& 0.966& $-0.217$& 1.012\\
HR1951&
52027.4415 & 0.10& 0.966&  --- &  --- & 0.135& 0.972& $-0.219$& 1.011\\
HR1966&
52029.3652 & 0.14& 0.949&  --- &  --- & 0.134& 0.976& $-0.113$& 1.001\\
HR2058&
52038.5645 & 0.11& 0.957&  --- &  --- & 0.178& 0.974& $-0.153$& 1.004\\
CCD1639&
52322.6262 & 0.08& 0.966&  --- &  --- & 0.158& 0.969& $-0.213$& 1.008\\
CCD1640&
52322.6737 & 0.05& 0.972&  --- &  --- & 0.165& 0.974& $-0.245$& 1.014\\
HR3339&
52343.4912 & 0.17& 0.961&  --- &  --- & 0.150& 0.965& $-0.480$& 1.039\\
HR3465&
52352.5557 & 0.15& 0.951&  --- &  --- & 0.153& 0.968& $-0.629$& 1.052\\
HR3492&
52362.3688 & 0.14& 0.962&  --- &  --- & 0.144& 0.972& $-0.314$& 1.018\\
HR3567&
52366.3847 & 0.08& 0.972&  --- &  --- & 0.122& 0.968& $-0.292$& 1.016\\
HR3658&
52373.4362 & 0.08& 0.970&  --- &  --- & 0.129& 0.975& $-0.365$& 1.025\\
HR3977&
52417.4422 & 0.17& 0.950&  --- &  --- & 0.139& 0.974& $-0.513$& 1.037\\
HR5335&
52657.5959 & 0.05& 0.981&  --- &  --- & 0.156& 0.972& $-0.330$& 1.020\\
HR5411&
52683.5671 & 0.09& 0.977&  --- &  --- & 0.140& 0.954& $-0.279$& 1.015\\
HR5431&
52684.4712 & 0.06& 0.976&  --- &  --- & 0.151& 0.969& $-0.341$& 1.023\\
HR5434&
52684.5418 & 0.06& 0.974&  --- &  --- & 0.134& 0.974& $-0.380$& 1.028\\
HR5481&
52687.4981 & 0.07& 0.977&  --- &  --- & 0.139& 0.975& $-0.383$& 1.026\\
HR5484&
52687.5881 & 0.07& 0.977&  --- &  --- & 0.130& 0.975& $-0.377$& 1.023\\
HR5495&
52688.4042 & 0.05& 0.979&  --- &  --- & 0.133& 0.969& $-0.373$& 1.024\\
HR5524&
52692.4424 & 0.06& 0.978&  --- &  --- & 0.151& 0.972& $-0.391$& 1.024\\
HR5537&
52693.4215 & 0.05& 0.978&  --- &  --- & 0.132& 0.970& $-0.389$& 1.022\\
\hline
\end{tabular}
\end{table*}

\addtocounter{table}{-1}
\begin{table*}
\caption{ -- continued}
\begin{tabular}{|l|l|cc|cc|cc|cc|} \hline \hline
\textsf{File} & \textsf{HJD}&
\multicolumn{2}{|c|}{\textsf{\ion{Si}{ii} 6347\,{\AA}}} &
\multicolumn{2}{|c|}{\textsf{\ion{Si}{ii} 6371\,{\AA}}} &
\multicolumn{2}{|c|}{\textsf{\ion{He}{i}  6678\,{\AA}}} &
\multicolumn{2}{|c|}{\textsf{\ion{Fe}{ii} 6456\,{\AA}}} \\
\cline {3-10}
& $-2400000$ & $W[\mbox{\AA}]$ & $I_c$
& $W[\mbox{\AA}]$ & $I_c$
& $W[\mbox{\AA}]$ & $I_c$
& $W[\mbox{\AA}]$ & $I_c$ \\
\hline
HR5757&
52720.317  & --- &  --- & 0.04 & 0.99 & 0.184& 0.967& $-0.486$& 1.033\\
HR5793&
52721.4772 & --- &  --- &  --- &  --- & 0.122& 0.968& $-0.323$& 1.020\\
HR5816&
52722.5142 & --- &  --- &  --- &  --- & 0.080& 0.976& $-0.481$& 1.036\\
HR5836&
52723.4036 & 0.05& 0.977&  --- &  --- & 0.136& 0.972& $-0.428$& 1.028\\
HR5837&
52723.4325 & --- &  --- &  --- &  --- & 0.152& 0.969& $-0.490$& 1.036\\
HR5864&
52724.4549 & --- &  --- &  --- &  --- & 0.101& 0.974& $-0.563$& 1.037\\
HR5926&
52727.4657 & 0.05& 0.971&  --- &  --- & 0.256& 0.954& $-0.380$& 1.022\\
md0412&
52734.7316 &0.063& 0.970&  --- &  --- & 0.178& 0.965& $-0.402$& 1.025\\
md1118&
52741.9524 &0.030& 0.980&  --- &  --- & 0.150& 0.971& $-0.385$& 1.025\\
md1221&
52742.6241 &0.067& 0.973&  --- &  --- & 0.145& 0.973& $-0.415$& 1.025\\
md1222&
52742.6307 &0.069& 0.964&  --- &  --- & 0.166& 0.969& $-0.468$& 1.036\\
md1421&
52744.8203 &0.076& 0.968&  --- &  --- & 0.134& 0.979& $-0.391$& 1.030\\
md1422&
52744.8238 &0.020& 0.981&  --- &  --- & 0.142& 0.969& $-0.393$& 1.019\\
md1426&
52744.8590 &0.030& 0.979&  --- &  --- & 0.145& 0.973& $-0.436$& 1.028\\
md2111&
52751.7782 &0.062& 0.972&  --- &  --- & 0.141& 0.977& $-0.380$& 1.028\\
md2112&
52751.7863 &0.066& 0.975&  --- &  --- & 0.158& 0.968& $-0.403$& 1.026\\
md2114&
52751.7979 &0.074& 0.975&  --- &  --- & 0.153& 0.973& $-0.400$& 1.024\\
md2115&
52751.8014 &0.026& 0.974&  --- &  --- & 0.161& 0.977& $-0.389$& 1.026\\
md2229&
52752.5113 &0.047& 0.973&  --- &  --- & 0.15 & 0.966& $-0.415$& 1.022\\
md2429&
52754.8454 &0.07 & 0.977&  --- &  --- & 0.157& 0.971& $-0.40 $& 1.025\\
md2430&
52754.8493 &0.032& 0.977&  --- &  --- & 0.107& 0.981& $-0.45 $& 1.025\\
md2431&
52754.8534 &0.053& 0.973&  --- &  --- & 0.133& 0.977& $-0.43 $& 1.029\\
md2432&
52754.8576 &0.043& 0.979&  --- &  --- & 0.15&  0.973& $-0.405$& 1.023\\
md2433&
52754.8600 &0.049& 0.97 &  --- &  --- & 0.147& 0.97 & $-0.406$& 1.03\\
md2434&
52754.8624 &0.064& 0.967&  --- &  --- & 0.133& 0.971& $-0.429$& 1.019\\
md2435&
52754.8644 &0.058& 0.977&  --- &  --- & 0.175& 0.97 & $-0.430$& 1.031\\
md2436&
52754.8680 &0.072& 0.966&  --- &  --- & 0.122& 0.982& $-0.4  $& 1.028\\
md2437&
52754.8699 &0.072& 0.971&  --- &  --- & 0.135& 0.971& $-0.431$& 1.025\\
md2438&
52754.8733 &0.043& 0.983&  --- &  --- & 0.141& 0.977& $-0.428$& 1.026\\
\hline
\end{tabular}
\end{table*}
\end{document}